\documentclass[epj]{svjour}
\usepackage{epsfig}
\usepackage{color}
\usepackage{amssymb}
\usepackage{amsmath, bbm}
\usepackage[figuresright]{rotating}

\usepackage[numbers,sort&compress]{natbib}
\usepackage{hyperref}

\newlength{\feynwidth} 

\newcommand{\La}{{\Lambda}}
\newcommand{\Si}{{\Sigma}}

\newcommand{\be}{\begin{eqnarray}}
\newcommand{\ee}{\end{eqnarray}}

\usepackage{graphics}
\usepackage{psfrag}
\usepackage{multirow,mathtools}
\usepackage{bbm}
\def\bea{\begin{eqnarray}}
\def\eea{\end{eqnarray}}
\def\beq{\begin{equation}}
\def\eeq{\end{equation}}

\newcommand{\Lc}{{\Lambda}_c}
\newcommand{\Sc}{{\Sigma}_c}
\newcommand{\Scs}{{\Sigma}_c^*}
\newcommand{\Yc}{{Y}_c}
\begin{document}
\title{Predictions for charmed nuclei based on $Y_c N$ forces inferred from 
lattice QCD simulations}
\titlerunning{Charmed nuclei}
\author{Johann Haidenbauer$^1$, Andreas Nogga$^1$, and Isaac Vida\~na$^2$}

\institute{
$^1$Institute for Advanced Simulation, Institut f\"{u}r Kernphysik (Theorie) and J\"{u}lich
Center for Hadron Physics, Forschungszentrum J\"{u}lich, D-52425 J\"{u}lich, Germany \\
$^2$Istituto Nazionale di Fisica Nucleare, Sezione di Catania. Dipartimento di Fisica
``Ettore Majorana", Universit\`a di Catania, Via Santa Sofia 64, I-95123 Catania, Italy
}
\date{Received: date / Revised version: date}

\abstract{
Charmed nuclei are investigated utilizing $\Lambda_c N$ and $\Sigma_c N$ interactions that have been extrapolated from lattice 
QCD simulations at unphysical masses of $m_\pi = 410$--$570$ MeV
to the physical point using chiral effective field theory as guideline. 
Calculations of the energies of $\Lambda_c$ 
single-particle bound states for various charmed nuclei from 
$^{\ 5}_{\Lambda_c}$Li to $^{209}_{\Lambda_c}$Bi are performed 
using a perturbative many-body approach. This approach allows one
to determine the finite nuclei $\Lambda_c$ self-energy 
from which the energies of the different bound states can be obtained.
Though the $\Lambda_c N$ interaction inferred from the lattice results 
is only moderately attractive, it supports the existence of charmed 
nuclei. Already the lightest nucleus considered is found to be bound.
The spin-orbit splitting of the p- and d-wave states turns out to be
small, as in the case of single $\Lambda$ hypernuclei. 
Additional calculations based on the Faddeev-Yakubovsky equations suggest 
that also $A=4$ systems involving a $\Lambda_c$ baryon are likely to be 
bound, but exclude a bound $^{\, 3}_{\Lambda_c}$He state. 
}

\maketitle

\section{Introduction} 
\label{sec:intro}

The prospect of an ample production of baryons with charm offered by 
facilities such as the LHC at CERN  
\cite{Acharya:2018,Aaij:2018,Sirunyan:2019},
RHIC at BNL \cite{Adam:2019},
J-PARC and KEK in Japan  
\cite{Noumi:2017,Niiyama:2018},
or FAIR in Germany 
\cite{PANDA,Wiedner:2011,Friman:2011}
has led to a renewed interest in the in-medium properties of such 
baryons \cite{Ohtani:2017,Carames:2018,Tsushima:2018,Yasui:2018} 
and also in the question whether they, and notably the lightest 
charmed baryon, the $\Lc$(2286), could form bound states with ordinary matter  
\cite{Liu:2012,Garcilazo:2015,Maeda:2016,vidana19,Wu:2020,Kopeliovich:2020}. 
In fact, there is a long history of speculations
about possible bound systems involving the $\Lc$  \cite{tyapkin75,Dover:1977,dover77b,iwao77,gatto78,bhamathi81,kolesnikov81,Bando:1982,bando83,Gibson:1983,Bando:1985,
bhamathi89,tsushima03b,tsushima04,tan04,Kopeliovich:2007} 
that started soon after the discovery of charmed baryons \cite{cazzoli75,knapp76} 
(see also the recent reviews \cite{Hosaka:2016,Krein:2017}). In 
principle, charmed nuclei could be produced by means of {\it charm exchange} 
or {\it associate charm production} reactions \cite{bressani89,bunyatov91}, 
in analogy to the ones widely used in hypernuclear physics. However, the experimental production of charmed nuclei is difficult due to the short lifetimes of $D$-meson beams which makes it necessary to place the target as close as possible to the $D$-meson production point, and due to the kinematics of the reactions: the charmed particles are formed with large momentum making their capture by a target-nucleus improbable. Because of these difficulties, 
up to now, only three albeit controversial candidates have been reported 
by an emulsion experiment carried out in Dubna in the mid-1970s \cite{batusov81a,batusov81b,batusov81d,lyukov89}.

The $\Lc N$ forces employed in the past investigations 
were predominantly derived within the meson-exchange framework, 
see Refs.~\cite{Liu:2012,vidana19} 
for recent examples, often utilizing SU(4) flavor symmetry in one 
form or the other. Lately, also the constituent quark model \cite{Garcilazo:2019} 
or a combination of meson-exchange and quark model \cite{Maeda:2016}
have been considered. Independently of that, in general, the resulting 
potentials turned out to be fairly attractive. 
Interestingly, a rather different picture emerged from recent lattice
QCD (LQCD) simulations by the HAL QCD collaboration \cite{miyamoto18,Miyamoto:2018}. 
Those studies, based on unphysical quark masses corresponding to pion masses 
of $m_\pi = 410 - 700$ MeV, suggest that the $\Lc N$ and $\Sc N$ interactions 
could be significantly less attractive than what had been proposed in the 
phenomenological studies mentioned above. 
In Ref.~\cite{Haidenbauer:2018}, an extrapolation of the HAL QCD results 
to the physical point was presented, using chiral effective 
field theory (EFT) \cite{Epelbaum:2003,Petschauer:2013} as guideline. 
It revealed that the $\Lc N$ interaction at the physical point is 
expected to be somewhat stronger than for large pion masses, however, 
still only moderately attractive and, specifically, considerably
less attractive than most of the phenomenological predictions for the 
$\Lc N$ interaction.

In the present work, we use the $\Lc N$ interaction from 
Ref.~\cite{Haidenbauer:2018} to explore the binding energies of 
charmed nuclei. In the literature different conventions for naming
$\Lc$ nuclei have been used in the past. 
We adopt here the standard nomenclature 
for nuclei with the proper generalization~\cite{Gibson:1983},  
spelled out for hypernuclei in Sect. I.B of Ref.~\cite{Gal:2016}. 
It takes into account that the characterizing letter(s) for the 
nucleus depends on its {total} charge {and not just} on the number of protons. 
For example, when adding the (uncharged) $\La$ to hydrogen ($^2$H) one
gets the hypertriton ($^3_{\La}$H) but adding the positively charged
$\Lc$ leads to $^{\ 3}_{\Lc}$He.
Similarly, based on this convention, the bound state of $\Lambda_c$ 
and $^{208}$Pb is $^{209}_{\Lc}$Bi. 

The lightest nuclei considered, {the $A=3$ and $A=4$ systems 
$^{\ 3}_{\Lc}$He and $^{\ 4}_{\Lc}$He }, are investigated by 
solving corresponding Faddeev-Yakubovsky equations.
Indeed, for the three-body system, bound states have been reported in 
Ref.~\cite{Maeda:2016} (with total angular momentum $J=1/2$ and $3/2$)
and Ref.~\cite{Garcilazo:2015} (for $J=3/2$). Note, however, that some of
the $\Lc N$ interactions employed in the former works are so strongly
attractive that they even predict two-body bound states (in the
$^1S_0$ as well as in the $^3S_1$ partial wave) with binding
energies comparable to that of the deuteron. 
For calculating heavier charmed nuclei, namely from $^{\ 5}_{\Lambda_c}$Li 
onward to $^{209}_{\Lambda_c}$Bi, a perturbative many-body 
approach is utilized that allows one to obtain the $\Lambda_c$ 
single-particle bound states in the 
different nuclei from the corresponding $\Lambda_c$ self-energy.
Results for charmed nuclei computed within this framework have been 
reported recently in Ref.~\cite{vidana19}, for a set of $Y_c N$ 
interactions deduced from an early version \cite{Reuber:1994} of the {hyperon-nucleon ($YN$)} 
meson-exchange potential of the J\"ulich Group \cite{Haidenbauer:2005} 
via SU(4) symmetry arguments. For these interactions, 
the $^{\ 5}_{\Lambda_c}$Li system 
{($^{\ 5}_{\Lambda_c}$He in the nomenclature used in \cite{vidana19})} turned out to be already bound. 
In this context, let us mention that the HAL QCD collaboration has 
likewise reported results for charmed nuclei~\cite{miyamoto18}. 
The calculations were performed with the $\Lc N$ potentials extracted 
from the lattice simulations 
at pion masses $410$-$700$ MeV, but using the physical masses of the $\Lc$ and 
the considered nuclei. Binding energies for $\Lc$ bound to 
$^{12}$C, $^{28}$Si, $^{40}$Ca, $^{58}$Na, $^{90}$Zr, and $^{208}$Pb were reported.  
 
Since after the publication of the $\Lc N$ interaction \cite{Haidenbauer:2018} new results from the HAL QCD collaboration became available that
include now the $\Sc N$ channel \cite{Miyamoto:2018}, we also 
revisit the $\Lc N$ interaction in order to explore in how far the
inclusion of a direct interaction in the $\Sc N$ channel modifies 
the extrapolation of the $\Lc N$ interaction from the results/masses 
of the lattice simulations to the physical point, and in how far
it influences the predictions for charmed nuclei. 
It turns out that adding/considering the interaction in the $\Sc N$ channel
has very little influence on the $\Lc N$ {scattering} results and also not on those for 
$\Lc$ nuclei. May be this is not too surprising in view of the fact that the 
thresholds of the two channels are separated by almost $170$ MeV. In any 
case, for completeness, we report predictions for the $\Sc N$ $S$-wave phase 
shifts based on our extrapolation of the HAL QCD results \cite{Miyamoto:2018}.

Another extrapolation of the lattice results for $\Sc N$, 
performed in heavy baryon chiral perturbation theory and 
taking into account heavy quark spin symmetry, has been
performed recently~\cite{Meng:2019}. Let us mention already now that 
neither in that work nor in our calculation any signal for resonances 
or bound states in the $\Sc N$ channel are found. 
The existence of such resonances has been suggested
in Ref.~\cite{Huang:2013}, where the $\Lc N$-$\Sc N$-$\Scs N$ 
interaction was investigated within the framework of the quark 
delocalization color screening model. 
Bound states (and resonances) in the $\Sc N$ $^3S_1$ partial wave 
around the $\Sc N$ and $\Scs N$ thresholds were also predicted
in Ref.~\cite{Maeda:2018} based on a $Y_c N$ interaction 
from meson exchange supplemented by short-range repulsion 
from a quark exchange model.

The paper is structured in the following way: A summary of the main characteristics of the $\Lambda_c N$ and $\Sigma_c N$ potentials is presented in Sec.~\ref{sec:formalism} . Results of the properties of $\Lambda_c$ in infinite nuclear matter, and light and heavier nuclei are reported in Sec.~\ref{sec:nuc_prop}. Finally, a brief summary and some concluding 
remarks are given in Sec.~\ref{sec:concl}.  The appendix summarizes results for $\Sc N$ scattering.


\section{The $\Lc N$ and $\Sc N$ potentials}
\label{sec:formalism}

The $\Lc N$ and $\Sc N$ interactions are constructed by using chiral EFT 
as guideline, following the scheme employed in our studies of the 
$\Lambda N$ and $\Sigma N$ systems 
\cite{Polinder,Haidenbauer:2013,Haidenbauer:2019,Petschauer:2020}. 
We summarize the essentials below. More details of the approach can be found in 
Ref.~\cite{Haidenbauer:2018}. 
The $Y_c N$ potential consists of contact terms and contributions from pion exchange.
The former are given by 
\begin{eqnarray}
V(^1S_0) = {\tilde{C}}_{^1S_0} + \tilde{D}_{^1S_0} m^2_\pi + ({C}_{^1S_0} + {D}_{^1S_0} m^2_\pi)\, ({p}^2+{p}'^2), &&\nonumber \\
V(^3S_1) = {\tilde{C}}_{^3S_1} + \tilde{D}_{^3S_1} m^2_\pi + ({C}_{^3S_1} + {D}_{^3S_1} m^2_\pi)\, ({p}^2+{p}'^2), &&\nonumber \\
V(^3D_1 -\, ^3S_1) = {C}_{\varepsilon_1}\, {p'}^2, \hskip 4.9cm && \nonumber \\
V(^3S_1 -\, ^3D_1) = {C}_{\varepsilon_1}\, {p}^2, \hskip 5.0cm &&  
\label{LEC}
\end{eqnarray}
for the partial waves considered in the present study. 
Here $p = |{\bf p}\,|$ and ${p}' = |{\bf p}\,'|$  are the initial and final 
center-of-mass (c.m.) momenta in the $\Lc N$ or $\Sc N$ systems. 
The quantities ${\tilde{C}}_{i}$, ${\tilde{D}}_{i}$, ${C}_{i}$, ${D}_{i}$ are 
low-energy constants (LECs) that are fixed by a fit to lattice data (phase shifts) by
the HAL QCD collaboration at $m_\pi=410$ MeV and $570$ MeV.
The $m_\pi$ dependence in Eq.~(\ref{LEC}) is motivated by the 
corresponding expression in the standard 
Weinberg counting up to next-to-leading order (NLO) \cite{Epelbaum:2003,Petschauer:2013} but 
differs from it by the term proportional 
to $m^2_\pi\, ({p}^2+{p}'^2)$ which is nominally of higher order. Nevertheless, 
we included that term in Ref.~\cite{Haidenbauer:2018} because it allowed us to obtain 
an optimal description of the LQCD results at $m_\pi=410$ MeV as well as $570$ MeV. 
We consider this as a prerequisite for a well constrained extrapolation to lower pion masses. 
Without such a term the phase shifts by HAL QCD for $m_\pi=410$ MeV would be underestimated
at low energies, as exemplified by results shown in Figs.~6 and 7 of Ref.~\cite{Meng:2019}. 

The contribution of pion exchange to the $\Yc N$ potential is given by 
\begin{equation}
V^{OPE}_{{Y_c N\to Y_c N}} =-f_{{Y_c Y_c\pi}}f_{{NN\pi}}
\frac{\left({\bf \sigma}_1\cdot {\bf q} \right)
\left({\bf \sigma}_2\cdot {\bf q} \right)}{ {\bf q}^{\,2}+m_{\pi}^2} \ ,
\label{OPE}
\end{equation}
where ${\bf q}$ is the transferred momentum, ${\bf q} = {\bf p'} - {\bf p}$.
The coupling constants $f_{BB'\pi}$ are related to the axial-vector strength
via $f_{BB'\pi} = g_A^{BB'}/2F_\pi$ with $F_\pi$ being the pion decay constant ($F_\pi\approx 93$~MeV).  
The coupling constant for the $\Lc\Sc\pi$ vertex can be 
determined from the experimentally known $\Sc \to \Lc\pi$ decay rate, see 
Refs.~\cite{Albertus:2005,Can:2016}. For the $\Sc\Sc\pi$ coupling constant 
lattice QCD results \cite{Alexandrou:2016} are employed. 
To be concrete, $g_A^{\Sc\Sc}=0.71$ \cite{Alexandrou:2016} and 
$g_A^{\Lc\Sc}=0.74$ \cite{Albertus:2005,Can:2016} are used, together
with $g_A^{NN}=1.27$ \cite{PDG}.
Note that $f_{\Lc\Lc\pi} \equiv 0$ under the assumption that isospin is
conserved. 
 
Besides the coupling constants at the physical point, one needs also 
their $m_\pi$ dependence:
\begin{equation}
f_{BB'\pi} (m^2_\pi) = \frac{g^{BB'}(m^2_\pi)}{2\,F_\pi (m^2_\pi) } \ .
\label{fpi}
\end{equation}
Results for the dependence of $F_\pi$ on $m^2_\pi$ are
available from lattice simulations \cite{Durr:2013}. 
Based on that work the values 
$F_\pi \approx 112$ MeV at $m_\pi = 410$ MeV and
$F_\pi \approx 129$ MeV at $m_\pi = 570$ MeV
were deduced and employed in Ref.~\cite{Haidenbauer:2018}.
With regard to the dependence of $g^{BB}_A$ on $m^2_\pi$ 
lattice QCD simulations indicate a rather small variation, 
at least for $g^{\Sc\Sc}_A$ and $g^{NN}_A$ where concrete
results are available \cite{Alexandrou:2014,Alexandrou:2016}. 
Because of that the dependence of the $g_A$'s on $m_\pi$ was neglected 
in Ref.~\cite{Haidenbauer:2018} and the values at the
physical point were used throughout. 

There was no information from LQCD on the $\Sc N$ interaction
at the time when the study in Ref.~\cite{Haidenbauer:2018} was performed,
and, thus, the interaction in the $\Sc N$ channel was not considered. 
Nonetheless, the coupling of $\Lc N$ to $\Sc N$ via pion exchange was already 
included. Due to its long-range nature, this coupling plays an important role 
in case of the $\Lambda N$ and $\Sigma N$ systems 
\cite{Haidenbauer:2013,Haidenbauer:2019}. 
Since $f_{\Lc\Sc\pi}$ is empirically known and only slightly smaller than 
$f_{\La\Si\pi}$ \cite{Haidenbauer:2013} the coupling between $\Lc N$ 
and $\Sc N$ via pion exchange should still be of relevance. 
Indeed, the effective contribution to the $\Lc N$ interaction,
$V_{\Lc N \to \Lc N} \sim V^{OPE}_{\Lc N \to \Sc N} \, G_{\Sc N}\, V^{OPE}_{\Sc N \to \Lc N}$, 
is expected to be smaller for energies around the $\Lc N$
threshold as compared to the situation for $\La N$, but just by a factor 
$2-3$. The reduction is due to the larger 
mass difference, $M_{\Sc}-M_{\Lc} \approx 167$ MeV versus $M_{\Sigma}-M_{\Lambda} \approx 78$ MeV,
that enters in the corresponding Green's functions 
$G_{\Sc N}$ or $G_{\Si N}$.
In any case, formally two-pion exchange contributions to the 
$\Lc N$ interaction involving the $\Sc$ do arise at NLO
\cite{Haidenbauer:2013}, and it can be expected that the piece with 
a $\Sc N$ intermediate state provides the dominant contribution \cite{Beane:2005}. 
In the actual calculation, this (reducible) two-pion exchange contribution 
to the $\Lc N$ potential is generated by solving a coupled-channel 
Lippmann-Schwinger {(LS) equation} (see below). 
Further (irreducible) NLO contributions from two-pion exchange \cite{Haidenbauer:2013} 
have been omitted in \cite{Haidenbauer:2018} for simplicity 
reasons. It was assumed that those can be effectively absorbed 
into the contact terms. 

For the present study, we add a direct $\Sc N$ interaction. The potential 
is determined in the same way as the one for $\Lc N$, by using results of
the HAL QCD collaboration for the corresponding $^3S_1$ phase shift at 
$m_\pi=410$ MeV and $570$ MeV \cite{Miyamoto:2018}. 
The main goal of this extension is to explore in how far the inclusion of a 
direct $\Sc N$ interaction modifies the $\Lc N$ results achieved earlier. 
Of particular interest is the question, whether it influences the results 
for charmed nuclei that we are concerned with here. 
Results for $\Sc N$ scattering itself are discussed and summarized in the appendix. 
 
The reaction amplitudes are obtained from the solution of a coupled-channel 
LS equation for the interaction potentials. 
After partial-wave projection \cite{Polinder}, the equation is given by
\begin{eqnarray}
T^{\nu'\nu,J}_{\rho'\rho}(p',p;\sqrt{s})&=&V^{\nu'\nu,J}_{\rho'\rho}(p',p) \nonumber\\
&+& \sum_{{\rho''},\nu''}\int_0^\infty \frac{dp''p''^2}{(2\pi)^3} \, 
V^{\nu'\nu'',J}_{\rho'\rho''}(p',p'') \nonumber \\
&\times& \frac{2\mu_{\rho''}}{p^2_{\rho''}-p''^2+i\eta}
T^{\nu''\nu,J}_{\rho''\rho}(p'',p;\sqrt{s}) \ ,
\label{LS}
\end{eqnarray}
where the label $\rho$ indicates the channels ($\Lc N$, $\Sc N$) and the label $\nu$ 
the partial wave. The quantity $\mu_\rho$ signifies the pertinent reduced mass. 
The on-shell momentum in the intermediate state, $p_{\rho}$, is defined by 
$\sqrt{s}=\sqrt{M^2_{B_{1,\rho}}+p_{\rho}^2}+\sqrt{M^2_{B_{2,\rho}}+p_{\rho}^2}$.

Since the integral in the LS equation (\ref{LS}) is divergent for the chiral 
potentials specified above, a regularization scheme needs to be 
introduced \cite{Epelbaum:2008,Machleidt:2011}.
For that purpose the potentials in the LS equation are cut off in 
momentum space by multiplication with a regulator function 
\cite{Polinder,Haidenbauer:2013},  
$f(p',p) = \exp\left[-\left(p'^4+p^4\right)/\Lambda^4\right]$.
In Ref.~\cite{Haidenbauer:2018}, cutoff values $\Lambda =500$--$600$ MeV were
employed, in line with the range that yielded the best results in NLO studies of 
the $\La N$ and $\Si N$ interactions \cite{Haidenbauer:2013,Haidenbauer:2019}.
The variations of the results with the cutoff, reflecting uncertainties due to 
the regularization, will be indicated by bands.

The baryon masses corresponding to the LQCD simulations at $m_\pi= 410$ and 
$570$~MeV are taken from Ref.~\cite{miyamoto18}.
For the calculation at the physical point, we use the masses 
$M_{N} = 938.92$ MeV,
$M_{\Lc} = 2286.5$ MeV, and $M_{\Sc} = 2455$ MeV. 

\section{$\Lc$ nuclei and matter properties}
\label{sec:nuc_prop}

In this section, we report results on the properties of the $\Lc$
in infinite nuclear matter and on charmed nuclei. 
In the pertinent calculations, we employ $Y_c N$ interactions 
extrapolated from results of lattice simulations by the HAL QCD 
collaboration \cite{miyamoto18,Miyamoto:2018} to the 
physical point. In particular, we use the 
$\Lc N$ potential from Ref.~\cite{Haidenbauer:2018}, where
there is no direct $\Sc N$ interaction, and the
two potentials $Y_c N$-A and $Y_c N$-B, introduced and described
in the appendix, which include a direct $\Sc N$ interaction.
 
\begin{figure}[t]
\begin{center}
\includegraphics[width=85mm]{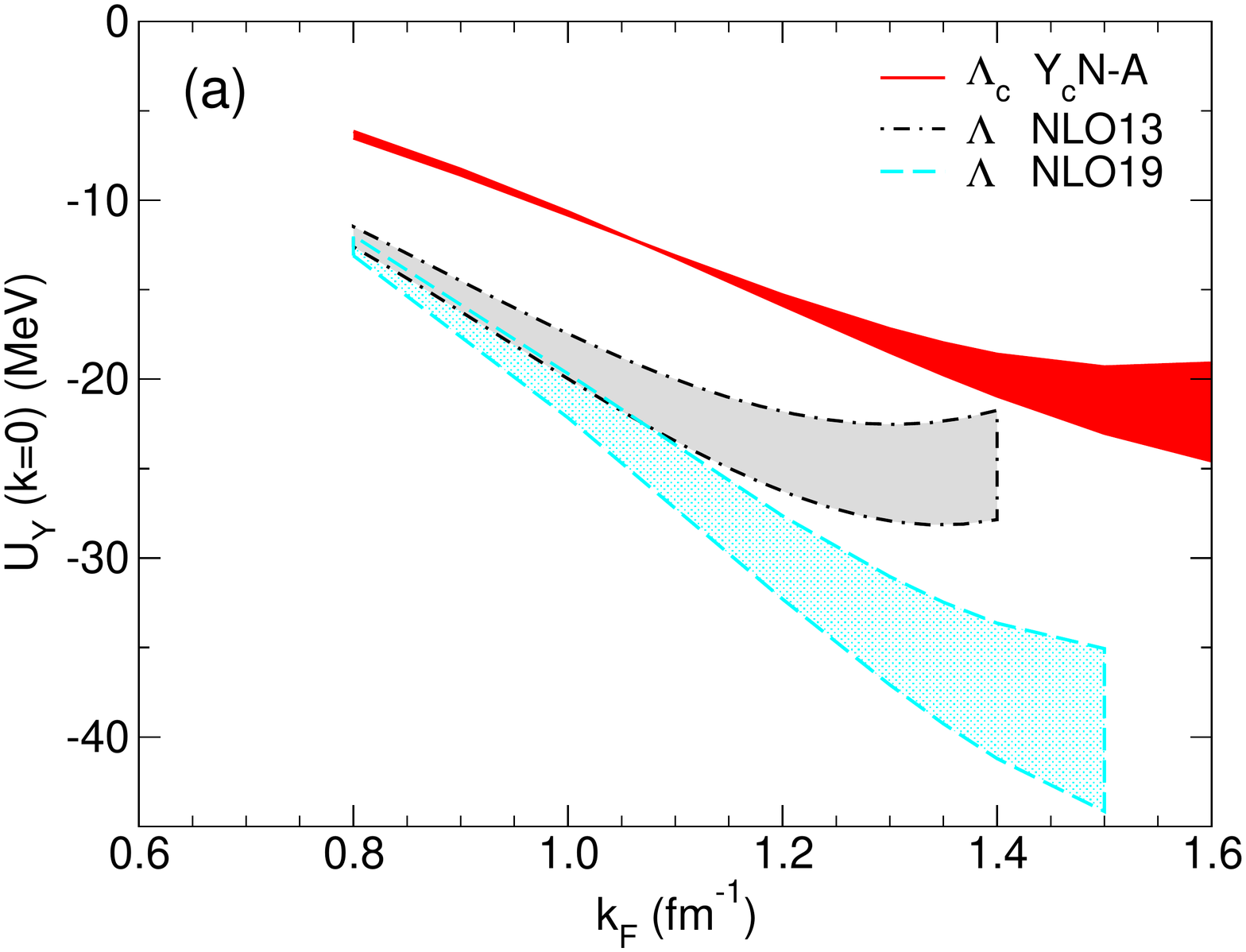}
\includegraphics[width=85mm]{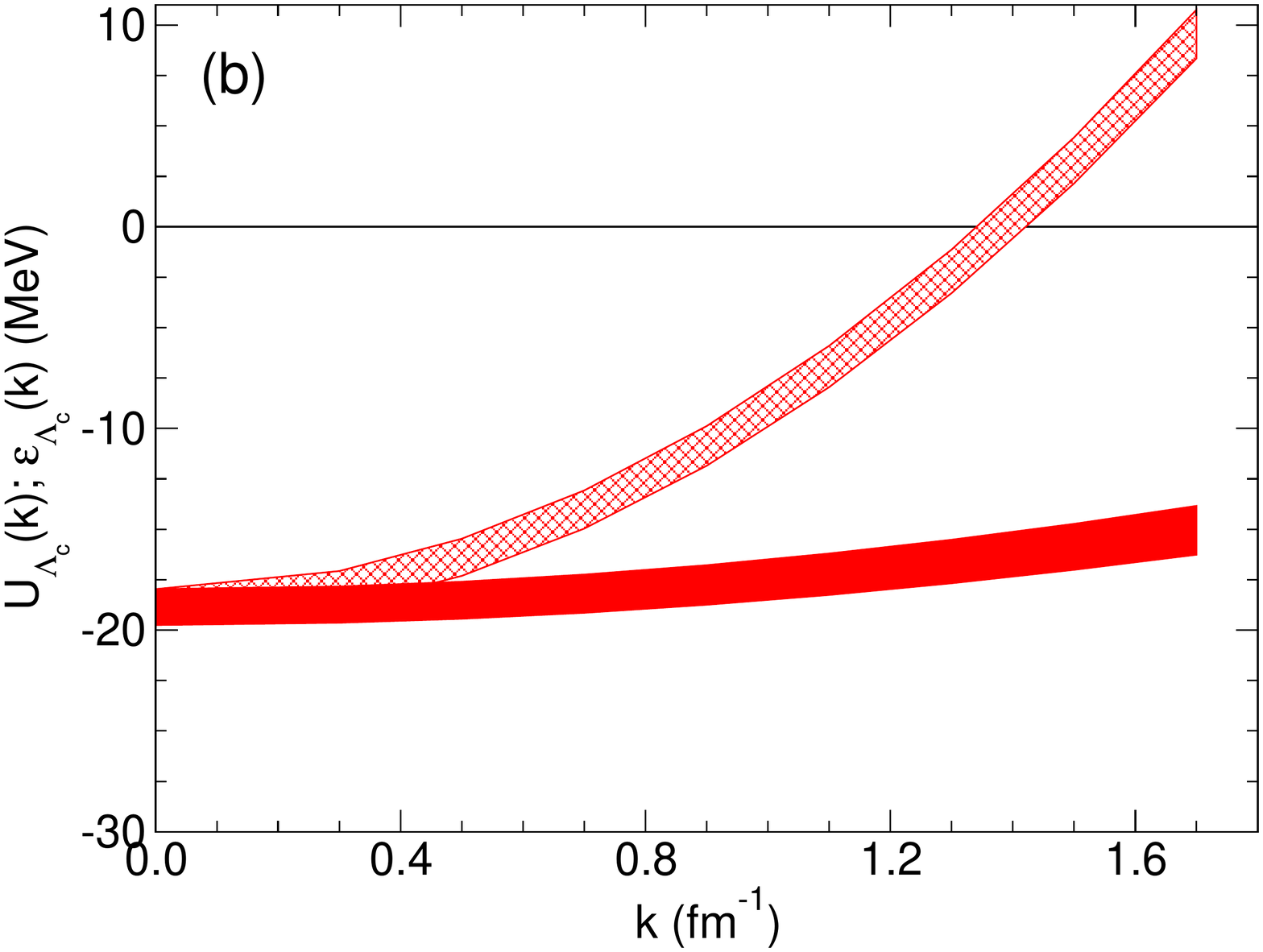}
\caption{$\Lc$ in symmetric nuclear matter. 
Top: The $\Lc$ s.p. potential $U_{\Lc}(k_{\Lc}=0)$ as a function of the
Fermi momentum $k_F$ in comparison to $U_{\Lambda}(k_{\Lambda}=0)$ 
\cite{Haidenbauer:2019} 
(NLO13 with dash-dotted border; NLO19 with dashed border) 
for cutoff variations $\La =500$-$600$ MeV. 
Bottom: Momentum dependence of $U_{\Lc}$ (solid band) and $\varepsilon_{\Lc}$ (hatched band) 
at the Fermi momentum $k_F=1.35$ fm$^{-1}$.
}
\label{gm}
\end{center}
\end{figure}

\subsection{$\Lc$ in infinite nuclear matter}

In order to investigate the properties of the $\Lc N$ interaction 
in nuclear matter, we perform a Brueckner--Hartree--Fock calculation
where we adopt the so-called discontinuous prescription when solving 
the Bethe--Goldstone equation. 
We follow closely our corresponding
calculation for the $\La N$ interaction \cite{Haidenbauer:2015}. In 
that work and similar ones (see, {\it e.g.}, Refs.~\cite{schulze97,vidana00}), 
the reader can find details how to solve the Bethe--Goldstone 
equation and how the single-particle (s.p.) potential $U_{\Lc}$ 
is determined self-consistently together with the G-matrices for a specific nuclear matter density $\rho$ (or 
Fermi momentum $k_F$). 

In Fig.~\ref{gm}(a), we present results for the dependence of $U_{\Lc}(k_{\Lc}=0)$
on the Fermi momentum, in comparison to those for the $\La$ hyperon obtained
with the NLO interaction from Refs.~\cite{Haidenbauer:2013,Haidenbauer:2019}.
In Fig.~\ref{gm}(b), we display the dependence of $U_{\Lc}(k_{\Lc})$ and 
the s.p. energy, $\varepsilon_{\Lc}(k_{\Lc})= k^2_{\Lc}/2m_{\Lc} + U_{\Lc}(k_{\Lc})$, 
on the $\Lc$ momentum at the Fermi momentum $k_F=1.35$ fm$^{-1}$, 
{\it i.e.,} at nuclear matter saturation density. 
The in-medium predictions for $\Lc$ are based on the $Y_c N$-A potential 
(cf. appendix) and the $\Lc N$ potential from Ref.~\cite{Haidenbauer:2018}. Results for the $\Lc$ properties for the alternative fit $Y_c N$-B, 
considered in the appendix, practically coincide with the ones for 
$Y_c N$-A and are therefore not shown. 

\begin{table*}[h]
\renewcommand{\arraystretch}{1.4}
\centering
\caption{$\Lc N$ scattering lengths (in fm) and partial-wave contributions 
to the s.p. potential $ U_{\Lc} (k_{\Lc} = 0)$ (in MeV)
at $k_F = 1.35 \ {\rm fm}^{-1}$.
Results are shown for the $Y_c N$-A potential which includes a direct 
$\Sc N$ interaction (cf. appendix), and for the $\Lc N$ 
interaction from Ref.~\cite{Haidenbauer:2018}. 
The cutoff values used ($\Lambda = 500,\,600$~MeV) are indicated in brackets. 
For comparison corresponding results for the $\Lambda N$ interactions
NLO13 \cite{Haidenbauer:2013} and NLO19 \cite{Haidenbauer:2019} are given. 
}
\vskip 0.2cm 
\begin{tabular}{ c|l||rr||rcrrrc|r }
\hline
\hline
& interaction & $a_{^1S_0}$ & $a_{^3S_1}$ & \multicolumn{7}{c }{$U_{\Lc}$ ($U_{\La}$)} \\
& &&& \ $^1S_0$ \ & \ $^3S_1+^3D_1$ \ & \ $^3P_0$ \ & \ $^1P_1$ \ & \ $^3P_1$ \ & \ $^3P_2+^3F_2$ \ & \ Total \\
\hline
$\Lc N$ & $\Lc N$ (500) \cite{Haidenbauer:2018} &-0.85&-0.81&  $-$ 5.1 &  $-$13.7 & $-$0.4 &    0.0 & $-$0.3 &   $-$0.2 &  $-$19.8  \\
        & $Y_c N$-A(500)                        &-0.85&-0.79&  $-$ 5.1 &  $-$13.5 & $-$0.4 &    0.0 & $-$0.3 &   $-$0.2 &  $-$19.7  \\
        & $\Lc N$ (600) \cite{Haidenbauer:2018} &-1.00&-0.98&  $-$ 5.5 &  $-$12.9 &    1.4 &    0.0 & $-$0.5 &   $-$0.4 &  $-$18.0  \\
        & $Y_c N$-A(600)                        &-1.00&-0.91&  $-$ 5.5 &  $-$12.4 &    1.4 &    0.0 & $-$0.5 &   $-$0.4 &  $-$17.6  \\
\hline
$\Lambda N$ & NLO13 (500) &-2.91&-1.61&  $-$15.3 &  $-$14.6 &    1.1 &    0.3 &  1.8 &   $-$1.3 &  $-$28.3  \\
            & NLO19 (500) &-2.91&-1.52&  $-$12.5 &  $-$28.0 &    1.1 &    0.3 &  1.8 &   $-$1.2 &  $-$39.3  \\
            & NLO13 (600) &-2.91&-1.54&  $-$12.3 &  $-$10.9 &    0.9 &    0.3 &  1.7 &   $-$1.1 &  $-$21.6  \\
            & NLO19 (600) &-2.91&-1.41&  $-$11.2 &  $-$22.8 &    0.9 &    0.4 &  1.7 &   $-$1.1 &  $-$32.6  \\
\hline
\hline
\end{tabular}
\label{tab:La}
\renewcommand{\arraystretch}{1.0}
\end{table*}

Partial-wave contributions to $ U_{\Lc} (k_{\Lc} = 0)$
at $k_F=1.35$ fm$^{-1}$ are listed in Table~\ref{tab:La}. 
Note that the contributions of the $P$ waves come solely 
from two-pion exchange involving the $\Sc N$ intermediate state. 
The total potential depth amounts 
to around {$-20\div-18$} MeV and is quite insensitive to whether a direct $\Sc N$ interaction is included or not.
As a reminder, the ``empirical'' value in case of the $\La$ hyperon 
is {$-30\div-27$} MeV \cite{Gal:2016}. 
Comparing the $\Lc$ results with the $\La$ case in detail, one can see 
that the contribution in the $^1S_0$ partial wave is reduced by roughly
a factor three. This is well in line with the corresponding interaction
strengths; the $\Lc N$ scattering length is also about a factor three 
smaller than that for $\La N$, see Table~\ref{tab:La}. 
For the $^3S_1$ partial wave the $\Lc N$ and $\La N$ contributions (for 
NLO13 \cite{Haidenbauer:2013}) are of comparable magnitude, despite of
the fact that the $\Lc N$ interaction is less attractive as reflected in 
the corresponding scattering lengths which is about 30-50\,\% smaller 
than that for $\La N$ scattering.  
Obviously, for $\Lc N$ the dispersive effects, which play an important 
role for the contribution of that partial wave in case of the 
$\La$ \cite{Haidenbauer:2019}, 
are smaller because the $\Lc N$-$\Sc N$ coupling is weaker due to a 
weaker transition potential and/or due to the larger threshold separation. 
Apparently, that reduced effect compensates for the somewhat less 
attractive $\Lc N$ interaction. 
When comparing with the results for the NLO interaction from 2019 \cite{Haidenbauer:2019},
where the $\La N$-$\Si N$ transition potential is noticeably weaker,
one sees a clear correlation between the smaller $^3S_1$ scattering 
length and the reduced contribution to $U_{\Lc}$ 
(cf. Table~\ref{tab:La}).

Finally, let us compare our nuclear matter results with other predictions 
for $U_{\Lc}$ found in the literature. 
Ref.~\cite{Bando:1985} contains some results for $U_{\Lc}$ based on an $\Yc N$ 
potential that is adapted from one of the $YN$ meson-exchange potentials by the
Nijmegen Group by imposing SU(4) flavor symmetry. In that work, a value of  
$ U_{\Lc} (k_{\Lc} = 0)\approx -25$ MeV at nuclear matter saturation density
has been found. However, note the large contributions from $P$ waves in that 
study. The two $S$ wave alone yield only around $-3.5$~MeV. 
A study utilizing parity-projected QCD sum rules \cite{Ohtani:2017}  
reports a potential depth of {$\sim -20$} MeV for $\Lc$ at nuclear matter 
saturation density. 
Yasui, in a perturbative approach based on a heavy-quark effective theory, 
finds a $\Lc$ binding energy of around {$-25\div-20$} MeV in nuclear matter 
\cite{Yasui:2018}. 
Though to some extent surprising, it is interesting to see that the achieved 
results are all fairly similar, despite of the different interactions
and approaches employed. 

\subsection{$^{\ 3}_{\Lc}$He and $^{\ 4}_{\Lc}$He systems}

For $A=3$ and $4$ charmed nuclei, Faddeev-Yakubovsky calculations
are performed in the same way as in former studies of hypernuclei
\cite{Miyagawa:1995,Nogga:2002,Haidenbauer:2019}.
As discussed in Ref.~\cite{Haidenbauer:2018}, 
based on the results for $\La$-hypernuclei and the relative strengths
of the $\Lc N$ and $\La N$ interactions, one can guess which light 
charmed nuclei could be bound. For that the pertinent mixtures of 
the spin-singlet and spin-triplet $\Lc N$ ($\La N$) interaction for 
$s$-shell nuclei are relevant \cite{Haidenbauer:2019,Gibson:1994}  
and, of course, the reduction of the kinetic energy associated with 
the $\Lc$ as a consequence of its larger mass \cite{Gibson:1983}. 
In view of the fact that, for the considered $Y_c N$ interactions,  
the $\Lc N$ $^1S_0$ scattering length is only one third of 
the one for $\La N$, while there is somewhat less difference in
the $^3S_1$ state (cf. Table~\ref{tab:La}), binding of light 
systems is expected to be mainly possible for charmed nuclei with a 
dominating spin-triplet $\Lc N$ contribution, i.e. $^{\, 3}_{\Lc}$He 
($J=\frac{3}{2}^+$), $^{\, 4}_{\Lc}$He ($1^+$), 
and $^{\, 5}_{\Lc}$Li \cite{Haidenbauer:2019,Miyagawa:1995,Gibson:1994}.  

Additionally, whereas the Coulomb interaction is of less importance
for $\Lambda$ separation energies of hypernuclei, its contribution to 
$\Lc$ separation energies of charmed nuclei is often decisive for binding 
\cite{Gibson:1983}. For the solution of the Faddeev-Yakubovsky 
equations here, we take the Coulomb interactions fully into account 
as described in \cite{Nogga:2000uu}. 

First, in order to benchmark our few-body calculations, we devised 
$\Lc N$ interactions that mimic the effective range parameters 
predicted by a $\Lc N$ potential obtained in the constituent quark 
model by Garcilazo et al.~\cite{Garcilazo:2019}. 
In that model, the triplet interaction is much stronger 
($a_{^3S_1} = -2.31$ fm) than the one in the singlet channel
($a_{^1S_0} = -0.86$ fm), cf. Table~3 in that reference.
Consequently, and in line with the above 
arguments and the explorations in Ref.~\cite{Miyagawa:1995}, 
a bound state for the $J=\frac{3}{2}$ state of $^{\ 3}_{\Lc}$He 
has been reported, with a $\Lc$ separation energy of approximately 
$140$~keV including Coulomb~\cite{Garcilazo:2015}.  
Since our interactions do not reproduce the phase shifts of the 
quark model perfectly over a larger energy region, we use 
two different realizations with cutoff $600$-$700$~MeV. We find 
separation energies between $60$~keV and $264$~keV. This includes 
a variation of $60$ keV due to different $NN$ interactions. 
The uncertainty due to different cutoffs of the $\Lc N$ 
interaction is larger than the one due to different $NN$ 
interactions. We also found that no bound state exists for that 
interaction for $^{\ 3}_{\Lc}$He with $J=\frac{1}{2}$.
This result confirms the earlier calculations of Garcilazo et al.

We then performed calculations for the $\Lc N$ potentials from
Ref.~\cite{Haidenbauer:2018}, and the interactions $Y_c N$-A 
and $Y_c N$-B of this work. Since all of these potentials predict 
a considerably weaker interaction in the $^3$S$_1$ partial wave, 
none of the charmed $A=3$ nuclei are found to be bound. 
This remains even true when the 
Coulomb interaction is not taken into account. 

Note that model 4 employed by Gibson et al.~\cite{Gibson:1983} 
has $\Lc N$ $^1S_0$ and $^3S_1$ scattering lengths close to
those of the $Y_c N$ interactions considered by us. No bound state 
for $A=3$ was found in that work, but a bound
$^{\, 4}_{\Lc}$He with a $\Lc$ separation energy of approximately 
$1.25$~MeV was predicted, after including an estimate 
for the contribution of the Coulomb interaction. 

\begin{table*}[tbp]
\renewcommand{\arraystretch}{1.4}
    \centering
    \caption{$\Lc$ separation energies $E_{\Lc}$ and binding energies with respect to breakup up into four baryons $E$ 
    for the $J=1^+$ state of $^{\ 4}_{\Lc}$He. The $NN$ interaction at order N$^4$LO+ with a cutoff of 450~MeV of Ref.~\cite{Reinert:2018} 
    was used leading to $E(^3{\rm H})=-8.141$~MeV. Expectation values for the kinetic energy $\langle T \rangle$, the 
    $NN$ potential energy $\langle V_{NN} \rangle$, and the $Y_c N$ potential energy $\langle V_{Y_c N} \rangle$ are also given. 
    The probability that the four-baryon have orbital angular momentum zero and two ( $P(S)$ and  $P(D)$ ) are 
    listed together with the probability $P(\Sc)$ for the $\Sc$ component. Energies are given in MeV and probabilities in \%.
    }
    \label{tab:4lamcHe1}
\begin{tabular}{l|r|r|rrrr|rrr}
\hline
\hline
interaction     &   $E_{\Lc}$ &     $E$ &   $\langle H \rangle$ & $\langle T \rangle$ & $\langle V_{NN} \rangle$ &  $\langle V_{Y_c N} \rangle$ & $P(S)$ &  $P(D)$ &  $P(\Sc)$ \\
\hline
    $\Lc N$(500) \cite{Haidenbauer:2018} & 0.13 &    -8.27 &   -8.26 &  36.82 &    -42.51 &      -2.58 &   93.21 & 6.75 & 0.21       \\
    $Y_c N$-A(500) & 0.11 &    -8.25 &   -8.25 &  36.65 &    -42.50 &      -2.39 &   93.22 & 6.73 & 0.19      \\
   $Y_c N$-B(500)& 0.11 &    -8.25 &   -8.25 &  36.65 &    -42.50 &      -2.39 &   93.22 & 6.73 & 0.19       \\
\hline
    $\Lc N$(600)  \cite{Haidenbauer:2018} & 0.37 &    -8.51 &   -8.50 &  39.71 &    -42.52 &      -5.69 &   92.81 & 7.13 & 0.46       \\
    $Y_c N$-A(600) & 0.30 &    -8.44 &   -8.43 &  39.15 &    -42.51 &      -5.07 &   92.88 & 7.06 & 0.39      \\
   $Y_c N$-B(600)& 0.30 &    -8.44 &   -8.43 &  39.15 &    -42.51 &      -5.07 &   92.88 & 7.06 & 0.39       \\
 \hline
 \hline
\end{tabular}
\renewcommand{\arraystretch}{1.0}
\end{table*}

\begin{table*}[tbp]
\renewcommand{\arraystretch}{1.4}
    \centering
    \caption{Same as Table~\ref{tab:4lamcHe1} for the $J=0^+$ state of $^{\ 4}_{\Lc}$He.}
    \label{tab:4lamcHe0}
\begin{tabular}{l|r|r|rrrr|rrr}
\hline
\hline
interaction     &   $E_{\Lc}$ &     $E$ &   $\langle H \rangle$ & $\langle T \rangle$ & $\langle V_{NN} \rangle$ &  $\langle V_{Y_c N} \rangle$ &  $P(S)$ &  $P(D)$&  $P(\Sc)$ \\
\hline
    $\Lc N$(500) \cite{Haidenbauer:2018} &   --  &    --  & -8.08 &   34.54 &    -42.51 &  -0.11 &   93.39 &  6.57 & 0.01       \\
    $Y_c N$-A(500) &   --  &    --  & -8.12 &   35.30 &    -42.55 & -0.86 &   93.33 &  6.63 & 0.09      \\
    $Y_c N$-B(500)&   --  &    --  & -8.12 &   35.30 &    -42.55 &  -0.86 &   93.33 &  6.63 & 0.09       \\
\hline    
    $\Lc N$(600) \cite{Haidenbauer:2018} & 0.18 & -8.32 & -8.30 &   39.21 &    -42.67 &      
    -4.84 &   92.99 &  6.95 & 0.47       \\
    $Y_c N$-A(600) & 0.10 & -8.25 & -8.23 &   38.17 &    -42.57 &      -3.83 &   93.09 &  6.85 & 0.34     \\
    $Y_c N$-B(600)& 0.10 & -8.25 & -8.23 &   38.17 &    -42.57 &      
    -3.83 &   93.09 &  6.85 & 0.34       \\
 \hline
 \hline
    \end{tabular}
    \renewcommand{\arraystretch}{1.0}
\end{table*}

The $A=4$ results for the interactions considered in this work are 
summarized in Tables~\ref{tab:4lamcHe1} and \ref{tab:4lamcHe0}. 
In all of our calculations, we found that the $J=1^+$ state is more 
bound than the $J=0^+$ state. As obvious from Table~\ref{tab:4lamcHe1}, 
the results are very independent on whether the direct $\Sc N$ interaction 
has been included or omitted. The cutoff of the $Y_c N$ interactions has 
a much larger effect on the energies than the inclusion of the direct interaction. Even more striking is the observation that the results for 
$A=4$ are identical for the potentials $Y_c N$-A and $Y_c N$-B. 
This shows unmistakably the insensitivity of the predicted bound-state 
properties on the $\Sc N$ channel. 
The separation energies for the $J=1^+$ state are between 
$100$~keV and $370$~keV, a clear evidence that $A=4$ charmed 
nuclei could be bound. 
The binding energies are, however, somewhat 
smaller than predicted in \cite{Gibson:1983}. 
We believe that this is partly due to omitting tensor interactions 
in the $Y_c N$ in \cite{Gibson:1983} which is fully taken into account 
in our calculations. 
For comparison, we have also used the interactions that simulate the quark-model potential of Ref. \cite{Garcilazo:2019}. This interaction 
clearly provides stronger binding 
leading to $1.2$ to $2.1$~MeV separation energy depending on the cutoff used. 

A few properties of the resulting wave functions are summarized in
Tables~\ref{tab:4lamcHe1} and \ref{tab:4lamcHe0}, too.
First of all, it is interesting to compare 
the expectation value of the Hamiltonian to the energy. 
For the numerical calculations, we need to restrict the 
number of partial waves. The most significant restriction 
is that the algebraic sum of all orbital angular momenta is less or equal 8. We checked that the solution of the Yakubovsky equations is converged such that the energy $E$ is accurate to approximately 10 keV. The expection values differ at most by 
20 keV. The slightly larger differences is due to a slower 
convergence of the wave functions compared to the Yakubovsky components. The good agreement of both numbers is a confirmation of the consistency of the numerical calculation. We also show separate expectation values of the kinetic 
energy, the $NN$ potential energy and the $Y_c N$ potential 
energy. It sticks out that the $NN$ potential energy is very similar for all considered bound states. Clearly, the 
nuclear core is not very much distorted by the presence of 
the charmed hyperon. The expectation value of the 
$Y_c N$ interaction is mainly dependent on the cutoff 
of the interaction and less dependent on the $\Sigma_c N$ 
contribution as can be seen from the similarity of the results of 
$\Lambda_c N$ \cite{Haidenbauer:2018}, $Y_c N$-A and $Y_c N$-B. 

Finally, we give probabilities for the total orbital
angular momentum of 0 and 2 $P(S)$ and $P(D)$. $P$-waves 
and $F$-waves only give a negligible contribution. Obviously, the tensor components of the $NN$ and $Y_c N$ interactions 
induce a $D$-wave contribution of approximately 7\%. 
The probability to find a $\Sigma_c$ is, similar to the 
$\Sigma$ in ordinary hypernuclei, small and depends strongly 
on the cutoff of the $Y_c N$ interaction. 

The outcome for the $J=0^+$ state is compiled in 
Table~\ref{tab:4lamcHe0}. In this case, we do not find 
a bound $^4_{\Lambda_c}$He for the interactions with a cutoff 
of 500~MeV. Such a state is however close to being bound as can 
be seen from the expectation values based on an approximate solution 
of the Yakubovsky equation, cf. Table~\ref{tab:4lamcHe0}.
Also for the larger cutoff, the $\Lambda_c$ separation energy is only $100$-$180$~keV. Therefore, for our interactions, we can neither confirm 
nor exclude the existence of a $0^+$ bound state. 
We note that we do find a bound $0^+$ state for the interactions that
simulate the quark model potential of Ref.~\cite{Garcilazo:2019}.
In that case the separation energy is between $130$~keV and $470$~keV 
depending on the cutoff used.  

Given that the dependence on the $NN$ interaction was smaller than the 
dependence on the employed regulator in the $Y_c N$ interaction for $A=3$, 
we refrain from repeating the computationally very expensive $A=4$ 
calculations for different $NN$ interactions. 
We do not expect that the results will be significantly different for other 
choices. 


\subsection{Heavier charmed nuclei}

\begin{table*}[ht!]
\caption{Energy of $\Lambda_c$ single-particle bound states (in MeV) of 
several charmed nuclei from $^{\ 5}_{\Lambda_c}$Li to 
$^{209}_{\Lambda_c}$Bi. For
convenience the corresponding core nucleus is indicated in brackets. 
Results are shown for two values of the cutoff $\Lambda=500,\, 600$ MeV.}
\renewcommand{\arraystretch}{1.6}
\begin{center}
\begin{tabular}{c|cc|cccc}
\hline
\hline
  & \multicolumn{2}{c|}{$\Lc N$ \cite{Haidenbauer:2018}} 
  &  \multicolumn{4}{c}{$Y_cN$} \cr
& (500) & (600) 
& A(500) & B(500) & A(600) & B(600)    \\
\hline
$^{\ 5}_{\Lc}$Li [$^{4}$He] & $$  & $$ & $$ & $$ & $$ & $$   \\ 
$1 s_{1/2}$ & $-0.86$  &  $-0.59$ &  $-0.82$ & $-0.81$ & $-0.47$ & $-0.45$  \\
\hline
$^{13}_{\Lc}$N [$^{12}$C] & $$  & $$ & $$ & $$ & $$ & $$   \\ 
$1 s_{1/2}$ & $-3.71$  &  $-2.78$ &  $-3.61$ & $-3.59$ & $-2.53$ & $-2.49$  \\ 
\hline
$^{17}_{\Lc}$F  [$^{16}$O] & $$  & $$ & $$ & $$ & $$ & $$   \\ 
 $1 s_{1/2}$ & $-5.18$  & $-3.70$ &  $-5.07$ & $-5.04$ & $-3.41$ & $-3.36$  \\ 
\hline
$^{41}_{\Lc}$Sc  [$^{40}$Ca]  & $$  & $$ & $$ & $$ & $$ & $$   \\ 
$1 s_{1/2}$ & $-6.36$  & $-4.35$ &  $-6.21$ & $-6.18$ & $-3.94$ & $-3.87$  \\ 
$1 p_{3/2}$ & $-0.93$  & $-$ &  $-0.79$ & $-0.77$ & $-$ & $-$  \\ 
$1 p_{1/2}$ & $-1.01$  & $-$ &  $-0.96$ & $-0.94$ & $-$ & $-$  \\ 
\hline
$^{91}_{\Lc}$Nb  [$^{90}$Zr] & $$  & $$ & $$ & $$ & $$ & $$   \\ 
$1 s_{1/2}$ & $-6.15$  & $-4.39$ &  $-5.95$ & $-5.91$ & $-3.88$ & $-3.79$  \\ 
$1 p_{3/2}$ & $-2.65$  & $-1.13$ &  $-2.46$ & $-2.42$ & $-0.65$ & $-0.57$  \\ 
$1 p_{1/2}$ & $-2.60$  & $-0.76$ &  $-2.41$ & $-2.37$ & $-0.29$ & $-0.21$  \\ 
\hline
$^{209}_{\Lc}$Bi  [$^{208}$Pb] & $$  & $$ & $$ & $$ & $$ & $$   \\ 
$1 s_{1/2}$ & $-4.89$  & $-2.15$ &  $-4.64$ & $-4.59$ & $-1.51$ & $-1.40$  \\ 
$1 p_{3/2}$ & $-4.37$  & $-1.73$ &  $-4.12$ & $-4.07$ & $-1.11$ & $-1.01$  \\ 
$1 p_{1/2}$ & $-4.36$  & $-1.52$ &  $-4.11$ & $-4.06$ & $-0.91$ & $-0.80$  \\ 
$1 d_{5/2}$ & $-0.18$  & $-$ &  $-$ & $-$ & $-$ & $-$  \\ 
$1 d_{3/2}$ & $-0.30$  & $-$ &  $-0.12$ & $-0.08$ & $-$ & $-$  \\ 
\hline
\hline
\end{tabular}
\end{center}
\label{tab:bound}
\renewcommand{\arraystretch}{1.0}
\end{table*}
Now we consider the energy of the $\Lambda_c$ single-particle bound states in heavier nuclei.  
To such end, we follow a perturbative many-body approach whose starting point is a nuclear matter $G$-matrix derived from the bare $Y_c N$ interactions described in Sec.~\ref{sec:formalism} and the appendix. 
This $G$-matrix is then used to calculate the self-energy 
of the $\Lambda_c$ in the finite nucleus. Solving the Schr\"{o}dinger equation with this self-energy, finally, we are able
to determine the energies of all the single-particle bound states of the $\Lambda_c$ in the nucleus. This approach also provides the 
real and imaginary parts of the $\Lambda_c$ optical potential at positive energies and, therefore, allows one to study
the $\Lambda_c$-nucleus  scattering properties. This method was already used to study the properties of the nucleon \cite{borromeo92}, 
the $\Delta$ isobar \cite{morten94} and the $\Lambda$ and $\Sigma$ hyperons \cite{morten96,vidana98,vidana17}
in finite nuclei, and very recently also those of the $\Lambda_c$ using a meson-exchange $Y_cN$ interaction \cite{vidana19}. 
A comprehensive description of the method can be found in these works and the interested reader is referred to any of them for details.

Results for  $^{\ 5}_{\Lc}$Li, $^{13}_{\Lc}$N , $^{17}_{\Lc}$F, $^{41}_{\Lc}$Sc, $^{91}_{\Lc}$Nb  
and $^{209}_{\Lc}$Bi are summarized in Table \ref{tab:bound} for the 
$\Lc N$ interaction from Ref.~\cite{Haidenbauer:2018} and the two 
potentials $Y_c N$-A and $Y_cN$-B with inclusion of a direct $\Sc N$
interaction. We note that all charmed nuclei considered consist of a 
closed-shell nuclear core plus a $\Lambda_c$ sitting in a single-particle state. We note also that, although the $\Lambda_cN$ interaction with the 
cutoff $600$ MeV is more attractive, the $\Lambda_c$ single-particle 
bound states predicted in this case are actually less bound. 
This is due to dispersive effects \cite{nogami70,bodmer71,dabrowski73} in 
the calculation of the $G$-matrix which suppress the contribution from 
the $\Lambda_cN\rightarrow\Sigma_cN\rightarrow\Lambda_cN$ coupling.
That coupling is significantly larger for the $600$ MeV cutoff and, accordingly, likewise the reduction of the overall attraction.
Before analyzing the results, we would like to point out that, as discussed in Ref.~\cite{Haidenbauer:2019V}, 
the approach followed tends to underestimate the energies of 
the $\Lambda$ hyperon single-particle bound states for light hypernuclei such as $^{\,5}_{\Lambda}$He.
Accordingly, we expect $^{\ 5}_{\Lc}$Li to be somewhat more strongly bound than what is suggested
by the values given in Table \ref{tab:bound}. 

It is interesting to observe that, contrary to single-$\Lambda$ hypernuclei where the $\Lambda$ is more and more bound when 
going from light to heavy nuclei,  the binding energy of the $\Lambda_c$ increases from $^{\ 5}_{\Lambda_c}$Li to $^{41}_{\Lambda_c}$Sc 
and then it decreases. This is due to the Coulomb repulsion between the $\Lambda_c$ and the protons of the nuclear core, which together 
with the kinetic energy of the $\Lambda_c$, compensates most of the attraction of the $\Lambda_cN$ interaction. The possible existence 
of $\Lambda_c$ nuclei is, therefore, subject to a delicate balance between the $\Lambda_cN$ interaction, the kinetic energy and the 
Coulomb force as it has been already pointed out in Refs.\ \cite{vidana19,tsushima03,tsushima03b,tsushima04,miyamoto18}. In particular, 
in Ref.\ \cite{miyamoto18} it was suggested that only light- or medium-mass $\Lambda_c$ nuclei could really exists whereas, for 
instance, in Ref.\ \cite{vidana19} it was found that even the heavier $\Lambda_c$ nucleus considered in that work, namely 
$^{209}_{\Lambda_c}$Bi, could exist, as in the present work. A small spin-orbit splitting of the $p-$ and $d-$wave states of the order 
of a few tenths of MeV is observed in all $\Lambda_c$ nuclei in agreement with the results obtained in 
Refs.\ \cite{vidana19,tsushima03,tsushima03b,tsushima04,tan04}. In addition, 
we note also that the level spacing of the $\Lambda_c$ 
single-particle states is smaller than those for the corresponding hypernuclei (see {\it e.g.} Table I of Ref.\ \cite{vidana17}). 
This is simply due to the fact that the mass of the $\Lambda_c$ is larger than that of the $\Lambda$ hyperon.

\begin{figure}[t]
\begin{center}
\vskip 1.2cm 
\includegraphics[width=85mm]{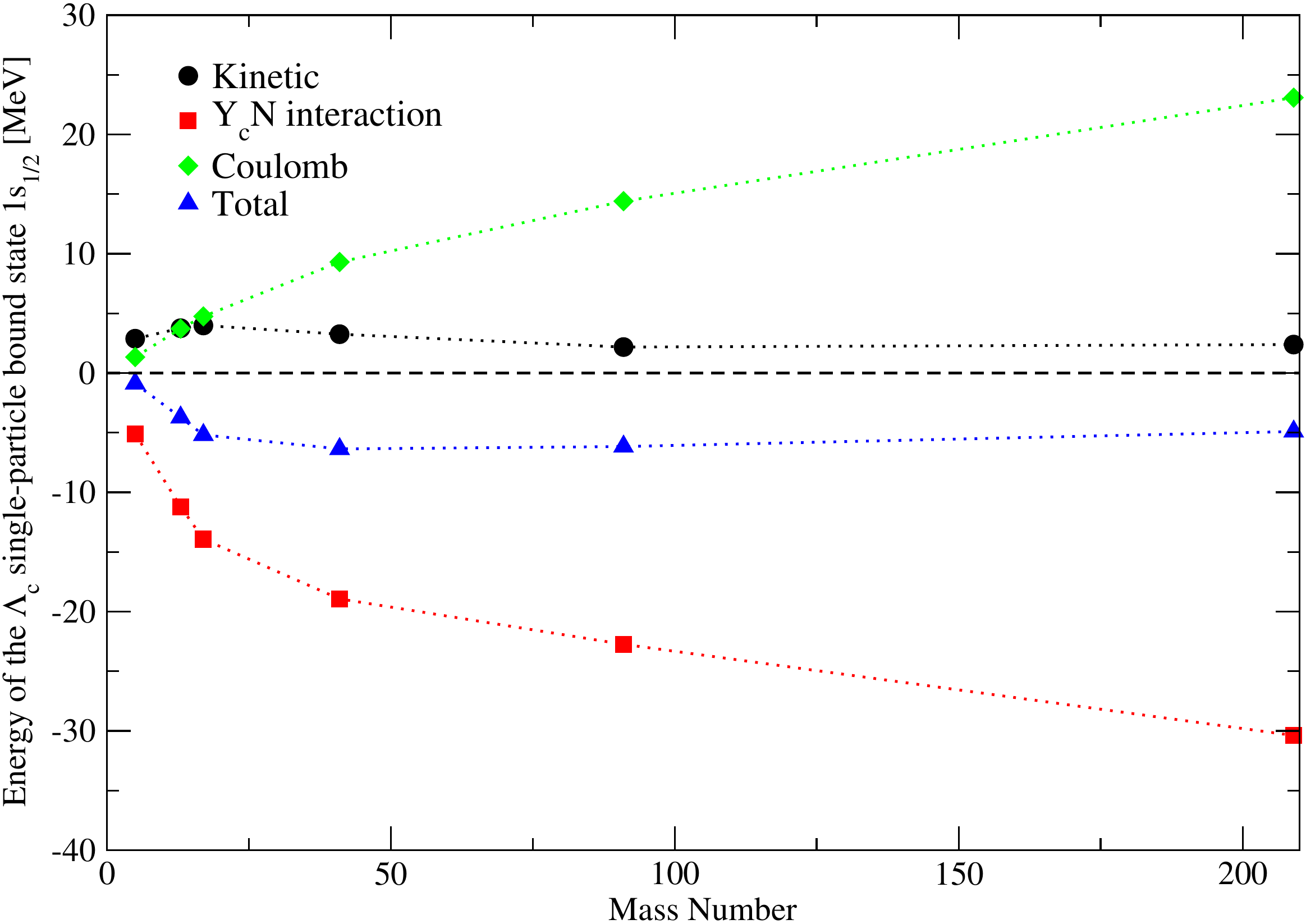}
\caption{Contributions of the kinetic energy, the $Y_cN$ interaction and the Coulomb potential to the energy 
of the $\Lambda_c$ single-particle bound state $1s_{1/2}$ as a function of the mass number of the $\Lambda_c$ nuclei 
considered. Results are shown for the potential $Y_c N$-A with a cutoff of $500$ MeV.}
\label{fig:coul}
\end{center}
\end{figure}

\begin{figure}[t]
\begin{center}
\includegraphics[width=85mm]{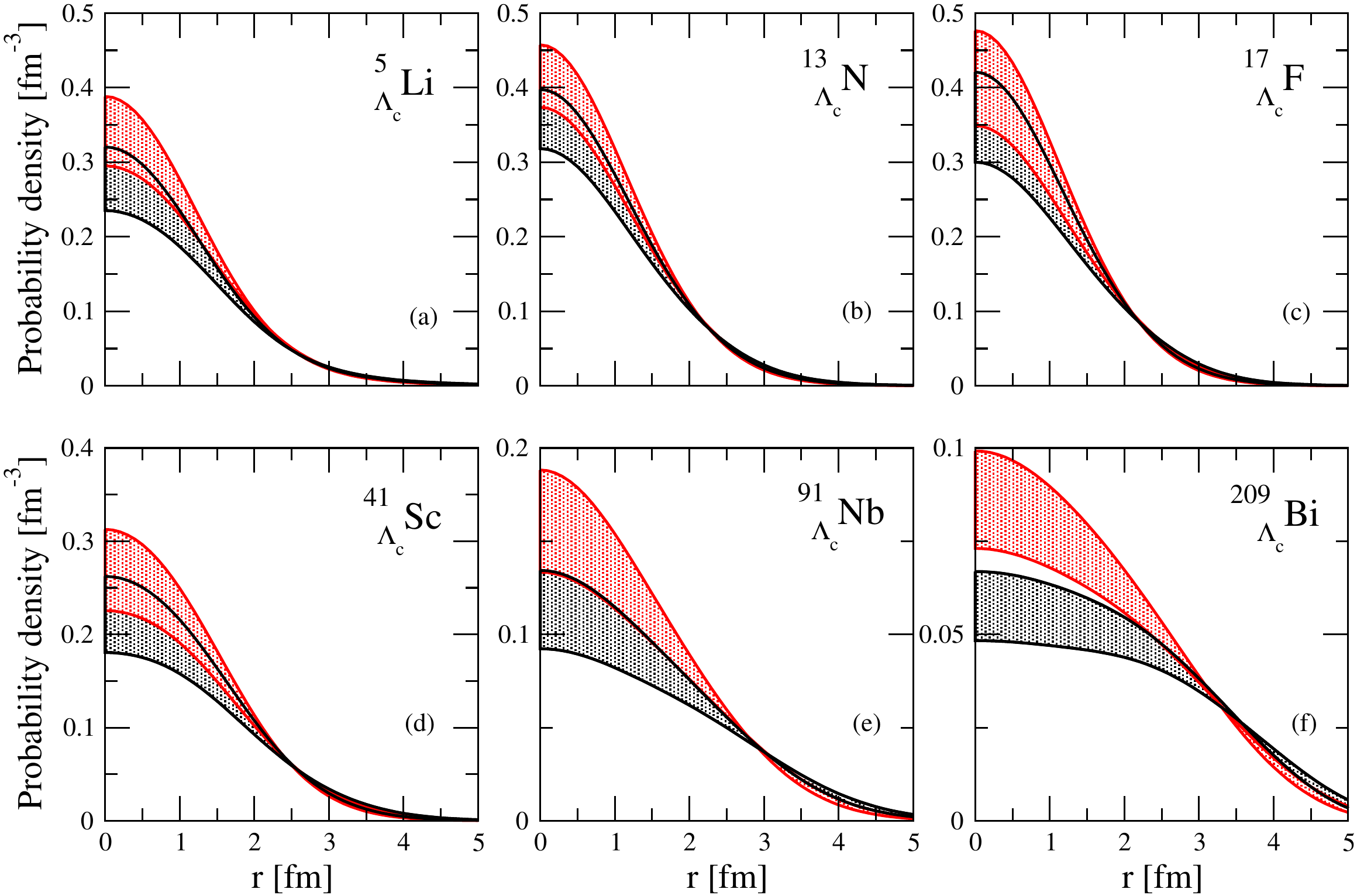}
\caption{$\Lc$ probability density distribution for the $1s_{1/2}$ state in the six $\Lc$ nuclei considered. Results are presented for the $Y_c N$-A potential. The dependence on the cut-off is indicated by bands. The red bands show the results when the Coulomb interaction is artificially switched off.}
\label{fig:wf}
\end{center}
\end{figure}

To understand better the role of the Coulomb force in our calculation, in Fig.\ \ref{fig:coul} we show the separate 
contributions of the kinetic energy, of the $Y_c N$ interaction, and of the Coulomb potential to the energy of the $\Lambda_c$ 
single-particle bound state $1s_{1/2}$ for the different charmed nuclei considered in this work as function of the mass number 
($A=N+Z$, with $N$ and $Z$ being the neutron and atomic numbers, 
respectively, of the specific nucleus). When going 
from light to heavy $\Lambda_c$ nuclei, the Coulomb contribution increases because of the increase of the atomic number 
whereas those of the kinetic energy and of the $Y_cN$ interaction decrease. The contribution of the kinetic energy decreases 
with the mass number because the wave function of the $1s_{1/2}$ state becomes more and more spread due to the larger 
extension of the nuclear density over which the $\Lambda_c$ wants to be distributed (see Fig.\ \ref{fig:wf}). The increase of the mass number leads 
to a more attractive $\Lambda_c$ self-energy (see, {\it e.g.,} Figs.\ 2 and 3 of Ref.\ \cite{vidana17} for a detailed 
discussion in the case of single-$\Lambda$ hypernuclei) that translates into a more negative contribution of the 
$Y_cN$ interaction.  Note that, when adding the three contributions they compensate in such a way that
the energy of the $1s_{1/2}$ decreases only by about 5 MeV from  $^{\ 5}_{\Lambda_c}$Li to $^{17}_{\Lambda_c}$F 
and then it increases very smoothly from $^{41}_{\Lambda_c}$Sc to $^{209}_{\Lambda_c}$Bi.

To end this section, we display in Fig.\ \ref{fig:wf} the probability density distribution ({\it i.e.,} the square of the radial wave function) of the $\Lambda_c$ in the $1s_{1/2}$ state for the six $\Lambda_c$ nuclei considered,
for the $Y_c N$-A. The cut-off dependence is indicated by bands. Results 
for the $\Lc N$ interaction from Ref.~\cite{Haidenbauer:2018} and
for $Y_c N$-B are not shown since the differences in the probability 
density are so small that they cannot be resolved in the plot.
Note that, when moving from light to heavy $\Lambda_c$ nuclei, due to the increase of the size of the nuclear core, the probability of finding the $\Lambda_c$ close to the center of the nucleus decreases, and it becomes more and more distributed over the whole nucleus.
The probability density distribution when the Coulomb interaction is artificially switched off is also shown for comparison.
Obviously, and as expected, the Coulomb repulsion pushes the $\Lambda_c$ away from the center of the nuclei. A similar effect is observed for the probability densities of the other $\Lambda_c$ single-particle bound states.


\section{Summary and Conclusions}
\label{sec:concl}

In the present work, we have investigated the binding energies 
of charmed nuclei. As input we used $\Lc N$ and $\Sc N$
interactions that have been extrapolated from lattice QCD simulations 
by the HAL QCD collaboration~\cite{miyamoto18,Miyamoto:2018} at
quark masses corresponding to $m_\pi=410 - 570$~MeV to the physical point. 
For this extrapolation, we used a framework based on chiral effective 
field theory~\cite{Haidenbauer:2018,Epelbaum:2003,Petschauer:2013}.
The $\Lc N$ interaction
established in this way is significantly weaker than what has 
been employed in most of the studies of charmed nuclei in the
literature so far.
The bound state calculations for light charmed nuclei have been 
carried out within the Faddeev-Yakubovsky framework. 
The results for heavier nuclei are from calculations 
of the energies of $\Lc$ single-particle bound states, 
performed within a perturbative many-body approach, which allows one
to determine the finite nuclei $\Lc$ self-energy from which 
the energies of the different bound states can be obtained.

Our results indicate that even for a weak $\Lc N$ interaction as 
suggested by the lattice simulations of the HAL QCD collaboration 
already $A=4$ charmed nuclei are likely to exist. 
Only the lightest nucleus considered, a charmed helium $^{\ 3}_{\Lc}$He, 
turned out to be unbound, in contrast to conjectures reported in 
Refs.~\cite{Maeda:2016,Garcilazo:2015}.

An additional aspect considered in the present work is the effect of
the $\Sc N$ interaction. Some results from lattice simulations for 
this channel have become available recently \cite{Miyamoto:2018}. 
There is admittedly a sizable uncertainty in the extrapolation 
of the HAL QCD results to the physical point, not least due to 
missing information on the behavior in the $\Sc^* N$ channel,
closely connected to the former by heavy quark spin symmetry. 
This makes reliable predictions for $\Sc N$ observables rather
difficult at the moment. 
On the other hand, we found that the uncertainties due to the present 
situation in the $\Sc N$ channel do not affect the conclusions 
on the properties of the $\Lc N$ interaction at low energies, relevant 
for the quest of charmed nuclei. Specifically, taking into account 
the coupling of $\Lc N$ to $\Sc N$ and the direct $\Sc N$ interaction 
as suggested by the HAL QCD results has very little influence on the 
existence of such bound states.
Indeed, for the $\Lc N$ interaction, the extrapolation of 
the lattice results to the physical point seems to be fairly reliable 
and stable and, therefore, we believe that robust predictions for the 
properties of the $\Lc$ in finite and infinite nuclear matter can be 
given based on the $\Lc N$ potentials established in 
Ref.~\cite{Haidenbauer:2018} and in this work. 

Prospects for detecting charmed nuclei at J-PARC have been discussed
at various occasions, see, e.g., Ref.~\cite{Tsunemi:2008}. 
Corresponding opportunities by the CBM experiment at FAIR are 
considered in Ref.~\cite{Steinheimer:2017}.  
The option for discovering charmed nuclei with neutrino beams is 
addressed in Ref.~\cite{Imai:2019}. 
An alternative on a different scope is offered by high-energy experiments 
such as the $pp$- and/or heavy-ion collisions \cite{Cho:2017}
presently pursued by the 
ALICE collaboration \cite{Braun:2018,Donigus:2019} at the LHC/CERN or 
the STAR collaboration at RHIC/BNL 
\cite{Agakishiev:2011,TheSTAR:2016}. 
Here ``exotic'' nuclei such as the anti-hypertriton or the 
$^4\overline{\rm He}$ were already produced and detected, and 
the lightest charmed nuclei might be within reach - now or in 
the near future - should they indeed exist. 
That said, one should be aware that there are tremendous experimental challenges for producing and detecting charmed nuclei, as has been summarized in Ref.~\cite{vidana19} but also indicated in the introduction to the present work. 

\begin{acknowledgement}
We acknowledge helpful communications with Vadim Baru, Benjamin D\"onigus, 
and Hirokazu Tamura.
This work is supported in part by the DFG and the NSFC through
funds provided to the Sino-German CRC 110 ``Symmetries and
the Emergence of Structure in QCD'' (DFG grant. no. TRR~110), and the COST Action CA16214.
The numerical calculations were performed on JURECA, the JURECA-Booster of the J\"ulich Supercomputing Centre, J\"ulich, Germany,
and the Centro di Calcolo of the INFN Sezione di Catania, Catania, Italy.
\end{acknowledgement}

\section*{Appendix: $\Sc N$ scattering}
\label{sec:results}

In this appendix, we discuss the results for $\Sc N$ scattering. However, let us 
emphasize from the beginning that these have to be interpreted with caution. 
For charmed baryons, heavy quark spin symmetry plays a role \cite{Meng:2019,Lu:2017}
and, thus, one should include not only the $\Sc N$ channel but also ${\Sc^*} N$. 
Indeed the $\Sc N$ and ${\Sc^*} N$ thresholds are just about $65$ MeV apart so 
that the coupling between those systems should be important. Unfortunately, 
there are no results for the ${\Sc^*} N$ interaction from lattice simulations 
and, therefore, it cannot be explicitly included in the analysis.
For that reason, it was omitted in our earlier study \cite{Haidenbauer:2018} and 
it was assumed that any effect of the ${\Sc^*} N$ interaction can be effectively 
absorbed into the $\Lc N$ LECs. After all, since $M_{\Sc^*}-M_{\Lc} = {234}$ MeV, 
there should be little influence on the low-energy $\Lc N$ amplitude anyway. 
In case of the $\Sc N$ channel, such an assumption is questionable.
In Ref.~\cite{Meng:2019} heavy quark spin symmetry was taken into account in
the derivation of the potential. But also in that work, the actual coupling of 
the $\Sc N$ and ${\Sc^*} N$ channels was ignored in the evaluation of the 
scattering amplitude. 

\begin{figure}[t]
\begin{center}
\includegraphics[width=85mm]{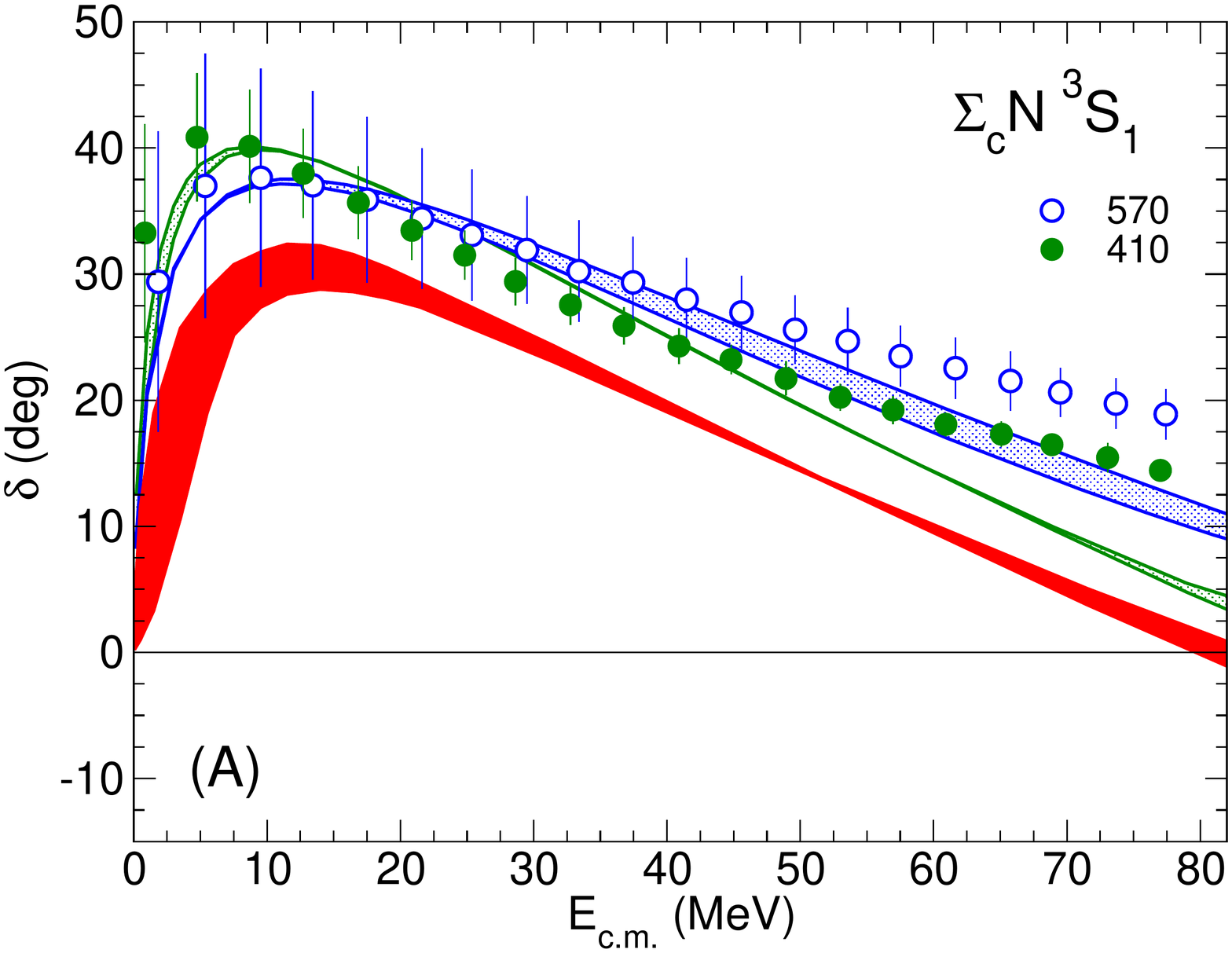}
\includegraphics[width=85mm]{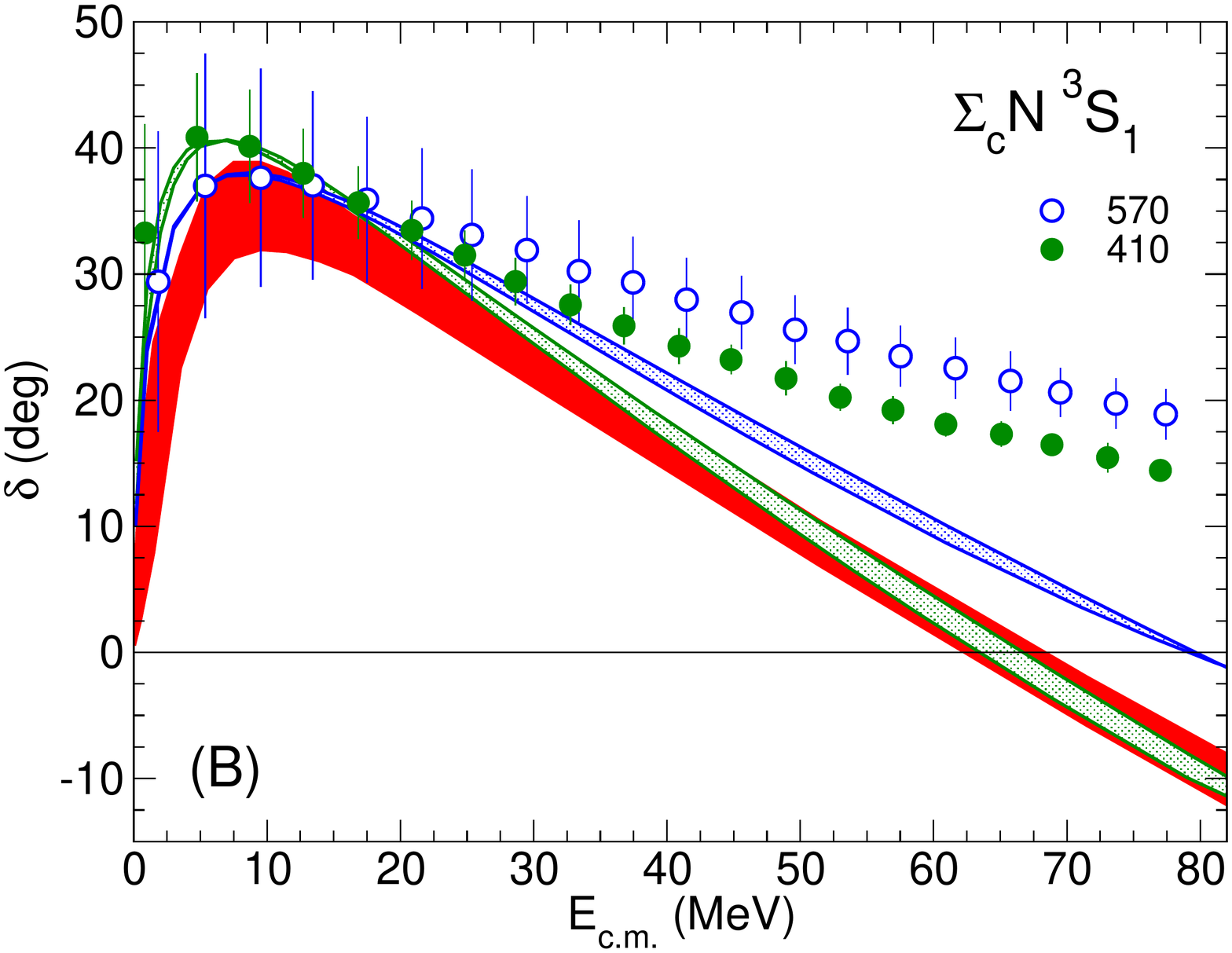}
\caption{$\Sc N$ $^3S_1$ phase shifts as function of the c.m. 
kinetic energy. 
Lattice QCD results~\cite{Miyamoto:2018} for $m_\pi=570$~MeV 
(open circles) and $410$~MeV (filled circles) are shown together with
our fits (blue and green narrow bands). 
The broader (red) band is the prediction for 
$m_\pi=138$~MeV. Two scenarios, $Y_c N$-A (top) and $Y_c N$-B (bottom)
are considered, see text. 
The bands represent the cutoff variation $\Lambda = 500$-$600$ MeV.
}
\label{ph3s1S} 
\end{center}
\end{figure}

Another complicacy comes from the fact that the mass difference
$M_{\Sc}-M_{\Lc}$ is larger than the pion mass. Because of that the use of
the static approximation for the pion exchange (see Eq.~(\ref{OPE})) in the 
$\Lc N \to \Sc N$ transition potential is rather problematic for 
energies near and/or above the $\Sc N$ threshold where the exchanged
pion can go on-shell. In principle, one should
take into account that there is a $\Lc N\pi$ three-body cut, see
Refs.~\cite{Baru:2011,Baru:2013,Baru:2015D} for a discussion on this issue for
the analogous $D\bar D^*$ - $D\bar D\pi$ case in the context of the $X(3872)$. 
As a matter of fact, for the unphysical pion masses of the LQCD calculation, 
this problem does not arise. However, it could have a noticeable influence on 
the extrapolation of the $\Sc N$ results to the physical point. 
This aspect is ignored in our calculation, and it is also not addressed 
in the $\Sc N$ study of Meng et al.~\cite{Meng:2019}. 
We emphasize that, for energies around the $\Lc N$ threshold, the 
static approximation is well justified. 

LQCD results for the $\Sc N$ $^3S_1$ phase shifts are available 
for $m_\pi = 410,\, 570,\, 700$ MeV \cite{Miyamoto:2018}. 
We determine the LECs of the contact interaction, cf. Eq.~(\ref{LEC}),
by a fit to the lattice data at the two lower pion masses. It
allows us to determine the LECs $\tilde C_i$ and $C_i$ as well
as $\tilde D_i$ and $D_i$, \textit{i.e.}, the ones which encode 
the pion-mass dependence of the contact interaction. 
The fits are done to the central values of the phase shifts as given 
in Fig.~2 of Ref.~\cite{Miyamoto:2018}, for energies 
up to $30$~MeV, cf. upper panel of Fig.~\ref{ph3s1S}. 
Alternative fits with particular emphasis on the near-threshold behavior 
of the HAL QCD results were performed, too, cf. the lower panel. 
Of course, in both cases, we made sure that we produce larger near-threshold
phase shifts for $m_\pi = 410$ MeV than for $570$ MeV, 
as suggested by the lattice simulation. 
It should be said that trying to fit to the HAL QCD results at somewhat
higher energies is not very meaningful
in view of the fact that the ${\Sc^*} N$ channel is not explicitly included.
Its threshold is around $85$~MeV for the HAL QCD calculations \cite{Miyamoto:2018}
and, as said above, at roughly $65$~MeV for physical masses. 

\begin{table}[t]
\renewcommand{\arraystretch}{1.5}
\centering
\caption{$\Lc N$ and $\Sc N$ ($I=1/2$) scattering lengths in 
the $^3S_1$ partial wave (in fm), for the $Y_c N$ 
potentials described in the text. In case of $\Sc N$ real and 
imaginary parts are given. $\Lc N$ results for the interaction 
from Ref.~\cite{Haidenbauer:2018} are listed as well. 
} 
\vskip 0.2cm
\begin{tabular}{l|c|c}
\hline
\hline
& $\Lc N$ & $\Sc N$ \\
\hline
$\Lc N$ (500) \cite{Haidenbauer:2018} & -0.81 & \\
$Y_c N$-A (500) & -0.79 & (-1.56, -1.35) \\
$Y_c N$-B (500) & -0.78 & (-2.09, -1.70) \\
$\Lc N$ (600) \cite{Haidenbauer:2018} & -0.98 & \\
$Y_c N$-A (600) & -0.91 & (-0.08, -1.97) \\
$Y_c N$-B (600) & -0.90 & (-0.21, -2.50) \\
\hline
\hline
\end{tabular}
\label{tab:ERE}
\renewcommand{\arraystretch}{1.0}
\end{table}

A combined fit to the $\Lc N$ and $\Sc N$ $^3S_1$ phase shifts turned out to 
be unnecessary because the inclusion of a direct $\Sc N$ interaction
had practically no effect on the $\Lc N$ results reported in \cite{Haidenbauer:2018}, 
at least for the energies considered there.  
We did not attempt to reproduce the inelasticity parameter,
given in Ref.~\cite{Miyamoto:2018} in terms of the $\Lc N$ $S$-matrix, 
$\eta = |S_{\Lc N,\Lc N}|$. In the lattice simulation it is with
values around $0.995$ basically compatible with $1$, which suggests that there
is practically no channel coupling. However, we believe that this could be an 
artifact of the way how the analysis by the HAL QCD collaboration is done. 
In the 
$\La N - \Si N$ systems, the strong channel coupling arises primarily from 
the tensor force mediated by pion exchange, and that leads to a strong 
coupling 
of the $^3S_1$ to the $^3D_1$ partial wave, see, e.g., Fig.~7 in 
Ref.~\cite{Haidenbauer:2013}. 
Indeed, for $^3S_1$$\to$$^3S_1$ transitions (as well as for $^1S_0$$\to$$^1S_0$), 
the expectation value of the tensor operator is zero. 
However, in the analysis of the HAL QCD collaboration, $D$ waves are not 
considered so that this component of the $Y_c N$ force is only effectively 
included in the $S$-wave interactions.   
In our calculation, we include the full one-pion exchange and that means that
there is a coupling to the $^3D_1$ partial wave - at the physical point
and also for the pion masses of the lattice simulation. 
Since the pion mass is known and the $\Lc\Sc\pi$ coupling constant is known, 
too (at least at the physical point), one can consider our result as genuine
prediction for the strength of the channel coupling and, thus, we did 
not impose any further constraints on it.  
 
Results for the $^3S_1$ phase shift are presented in Fig.~\ref{ph3s1S}
for pion masses of $570$~MeV and $410$~MeV, together with the extrapolation to 
$138$~MeV. The bands represent the dependence of the results on 
variations of the cutoff $\Lambda$.
One can see that the lattice results at $m_\pi=410$~MeV and $570$~MeV are 
reproduced quantitatively by our potential ($Y_c N$-A) up 
to c.m. energies of around $50$~MeV, cf. upper panel of Fig.~\ref{ph3s1S}. 
If we require a quantitative reproduction of the low-energy behavior 
($Y_c N$-B; lower panel), then there is 
agreement with the lattice results only up to around $25$~MeV. 
The cutoff dependence of the fits to the HAL QCD results is fairly small 
so that the (hatched) bands are barely visible. The phase shift 
obtained from the interaction at the physical point (solid bands) 
do exhibit a noticeable but still moderate cutoff dependence. 
That said, there is a sizable uncertainty in the extrapolation of the two 
fitting scenarios considered. When emphasis is put on the very low-energy
results by HAL QCD, then the phase shifts at the physical point are larger
and actually close to the lattice results for unphysical pion masses  
(Fig.~\ref{ph3s1S}; lower panel) whereas the fit over a larger energy 
region yields perceptibly smaller results. 
 
\begin{figure}[t]
\begin{center}
\includegraphics[width=85mm]{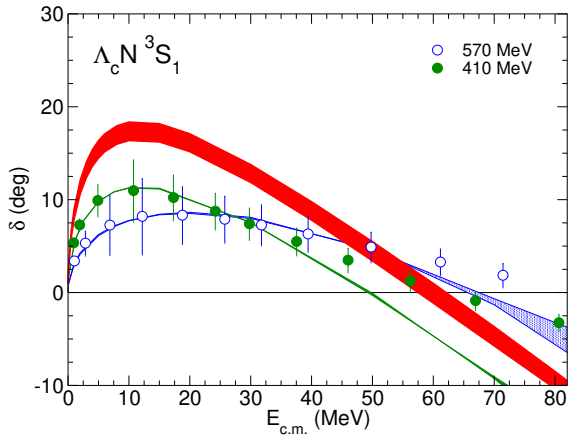}
\caption{Results for the $\Lc N$ $^3S_1$ phase shift for the interaction
$Y_c N$-A. 
Same description of curves and symbols as in Fig.~\ref{ph3s1S}. 
Lattice QCD results are from Ref.~\cite{miyamoto18}. 
}
\label{ph3s1L} 
\end{center}
\end{figure}

For convenience, we compiled the scattering lengths for $Y_c N$-A and
B in Table.~\ref{tab:ERE}.
Obviously, with the $\Sc N$ interaction included there is a small 
reduction in the $\Lc N$ $^3S_1$ scattering length (for A and for B)
as conpared to the results from Ref.~\cite{Haidenbauer:2018}.
However, the variation is rather moderate and stays within the 
uncertainty due to the regulator dependence. Also the variation in the 
corresponding $\Lc N$ $^3S_1$ phase shift is very small. Actually, 
the $\Lc N$ results with inclusion of a direct $\Sc N$ interaction, 
cf. Fig.~\ref{ph3s1L} for $Y_c N$-A, can be hardly distinguished from 
those presented in Fig.~2 in Ref.~\cite{Haidenbauer:2018}.
For the $\Sc N$ channel the situation is rather different.
First there is a sizable difference in the scattering length obtained
for the interactions A and B. In addition, and more disturbing, there 
is a fairly drastic regulator dependence. One can certainly
say that there is no indication for a near-by $\Sc N$ bound state,
a conclusion already drawn in Ref.~\cite{Meng:2019}. 
(The existence of $\Sc N$ bound states or resonances has been suggested 
in some studies in the past \cite{Huang:2013,Maeda:2018}). 
On the other hand, more quantitative conclusions are difficult to draw. 
Definitely, there is a significant coupling to the $\Lc N$ channel, 
mediated by one-pion exchange, which gives rise to an appreciable imaginary 
part of the scattering length in our calculation. Moreover, one should not 
forget that the static approximation is used by us for
simplicity reasons. In the real world another channel is open, 
namely $\Lc N\pi$, which contributes likewise to the inelasticity. 
An appropriate treatment is desirable but technically demanding and, 
thus, postponed to the future when hopefully more information from lattice 
simulations will be available. 
Fortunately, these uncertainties have basically no effect on
the predictions for the properties of the $\Lc N$ interaction at 
low energies, cf. Table.~\ref{tab:ERE} and Fig.~\ref{ph3s1L}, 
and also not on the binding energies of $\Lc$ nuclei, as discussed 
in the main part of this paper. 



\begin{thebibliography}{20}

\bibitem{Acharya:2018} 
  S.~Acharya {\it et al.} [ALICE collaboration],
  Phys.\ Lett.\ B {\bf 793}, 212 (2019).

\bibitem{Aaij:2018} 
  R.~Aaij {\it et al.} [LHCb collaboration],
  JHEP {\bf 1902}, 102 (2019). 

\bibitem{Sirunyan:2019} 
  A.~M.~Sirunyan {\it et al.} [CMS collaboration],
  arXiv:1906.03322 [hep-ex].

\bibitem{Adam:2019} 
  J.~Adam {\it et al.} [STAR collaboration],
  arXiv:1910.14628 [nucl-ex].

\bibitem{Noumi:2017} 
  H.~Noumi,
  JPS Conf.\ Proc.\  {\bf 17}, 111003 (2017).

\bibitem{Niiyama:2018} 
  M.~Niiyama {\it et al.} [Belle collaboration],
  Phys.\ Rev.\ D {\bf 97}, 072005 (2018). 

\bibitem{PANDA} 
  W.~Erni et al. [PANDA collaboration], 
  arXiv:0903.3905 [hep-ex].

\bibitem{Wiedner:2011} U.~Wiedner,
  Prog.\ Part.\ Nucl.\ Phys.\  {\bf 66}, 477 (2011). 
  
\bibitem{Friman:2011} 
  B.~Friman, C.~H\"ohne, J.~Knoll, S.~Leupold, J.~Randrup, R.~Rapp and P.~Senger,
  Lect.\ Notes Phys.\  {\bf 814}, pp. 1 (2011).

\bibitem{Ohtani:2017} 
  K.~Ohtani, K.~j.~Araki and M.~Oka,
  Phys.\ Rev.\ C {\bf 96}, 055208 (2017). 

\bibitem{Carames:2018} 
  T.~F.~Caram\'es, C.~E.~Fontoura, G.~Krein, J.~Vijande and A.~Valcarce,
  Phys.\ Rev.\ D {\bf 98}, 114019 (2018). 

\bibitem{Tsushima:2018} 
  K.~Tsushima,
  Phys.\ Rev.\ D {\bf 99}, 014026 (2019). 

\bibitem{Yasui:2018} 
  S.~Yasui,
  Phys.\ Rev.\ C {\bf 100}, 065201 (2019). 

\bibitem{Liu:2012} 
  Y.~R.~Liu and M.~Oka,
  Phys.\ Rev.\ D {\bf 85}, 014015 (2012).

\bibitem{Garcilazo:2015} 
  H.~Garcilazo, A.~Valcarce and T.~F.~Caram\'es,
  Phys.\ Rev.\ C {\bf 92}, 024006 (2015).

\bibitem{Maeda:2016} 
  S.~Maeda, M.~Oka, A.~Yokota, E.~Hiyama and Y.~R.~Liu,
  PTEP {\bf 2016}, no. 2, 023D02 (2016).

\bibitem{vidana19}
  I.~Vida\~na, A.~Ramos and C.~E.~Jim\'enez-Tejero,
  Phys.\ Rev.\ C {\bf 99}, 045208 (2019).

\bibitem{Kopeliovich:2020} 
  V.~B.~Kopeliovich and D.~E.~Lanskoy,
  arXiv:2001.04140 [nucl-th].

\bibitem{Wu:2020} 
  L.~Wu, J.~Hu and H.~Shen,
  Phys. Rev. C {\bf 101}, 024303 (2020).
  
\bibitem{tyapkin75} A. A. Tyapkin, Yad. Fiz. {\bf 22}, 181 (1975); Sov. J. Nucl. Phys. {\bf 22}, 89 (1976).

\bibitem{Dover:1977} 
  C.~B.~Dover and S.~H.~Kahana,
  Phys.\ Rev.\ Lett.\  {\bf 39}, 1506 (1977).

\bibitem{dover77b} C. B. Dover, S. H. Kahana and T. L. Trueman, Phys. Rev. D. {\bf 16}, 799 (1977).

\bibitem{iwao77} S. Iwao, Lett. Nuovo Cimento {\bf 19}, 647 (1977).

\bibitem{gatto78} R. Gatto and F. Paccanoni, Nuovo Cimento {\bf 46A}, 313 (1978).

\bibitem{bhamathi81} G. Bhamathi, Phys. Rev. C {\bf 24}, 1816 (1981).

\bibitem{kolesnikov81} N. N. Kolesnikov {\it et al.}, 
  Sov. J. Nucl. Phys. {\bf 34}, 957 (1981).

\bibitem{Bando:1982} 
  H.~Band\={o} and M.~Bando,
  Phys.\ Lett.\  {\bf 109B}, 164 (1982).

\bibitem{bando83} H. Band\={o} and S. Nagata, 
  Prog. Theor. Phys. {\bf 69}, 557 (1983).

\bibitem{Gibson:1983} 
  B.~F.~Gibson, C.~B.~Dover, G.~Bhamathi and D.~R.~Lehman,
  Phys.\ Rev.\ C {\bf 27}, 2085 (1983).

\bibitem{Bando:1985} 
  H.~Band\={o},
  Prog.\ Theor.\ Phys.\ Suppl. {\bf 81}, 197 (1985).

\bibitem{bhamathi89} G. Bhamathi, Nuovo Cimento A {\bf 102}, 607 (1989).

\bibitem{tsushima03b} K. Tsushima and F.~C. Khanna, Phys. Rev. C {\bf 67}, 015211 (2003).

\bibitem{tsushima04} 
  K.~Tsushima and F.~C.~Khanna,
  J.\ Phys.\ G {\bf 30}, 1765 (2004).

\bibitem{tan04} Y.-H. Tan and P.-Z. Ning, 
  Europhys. Lett. {\bf 67}, 355 (2004).

\bibitem{Kopeliovich:2007} 
  V.~B.~Kopeliovich and A.~M.~Shunderuk,
  Eur.\ Phys.\ J.\ A {\bf 33}, 277 (2007). 

\bibitem{cazzoli75} E. G. Cazzoli {\it et al.,} Phys. Rev. Lett.  {\bf 34}, 1125 (1975).

\bibitem{knapp76} B. Knapp {\it et al.,} Phys, Rev. Lett.  {\bf 37}, 882 (1976).

\bibitem{Hosaka:2016} 
  A.~Hosaka, T.~Hyodo, K.~Sudoh, Y.~Yamaguchi and S.~Yasui,
  Prog.\ Part.\ Nucl.\ Phys.\  {\bf 96}, 88 (2017). 

\bibitem{Krein:2017} 
  G.~Krein, A.~W.~Thomas and K.~Tsushima,
  Prog.\ Part.\ Nucl.\ Phys.\  {\bf 100}, 161 (2018).


\bibitem{bressani89} T. Bressani and F. Iazzi, Nuovo Cimento A {\bf 102}, 597 (1989).

\bibitem{bunyatov91} S. A. Bunyatov, V. V. Lyukov, N. I. Starkov and V. A. Tsarev, Nuovo Cimento A {\bf 104}, 1361 (1991).


\bibitem{batusov81a} Y. A. Batusov {\it et al.,} JINR Preprint E1-10069 (Dubna 1976).

\bibitem{batusov81b} Y. A. Batusov {\it et al.,} Pis'ma \v{Z}. \.{E}ksp. Teor. Fiz. {\bf 33}, 56 (1981); JETP Lett. {\bf 33}, 52 (1981).

\bibitem{batusov81d} Y. A. Batusov {\it et al.,} JINR Communication P1-85-495 (Dubna 1985).

\bibitem{lyukov89} V. V. Lyukov, Nuovo Cimento A {\bf 102}, 583 (1989).


\bibitem{Garcilazo:2019} 
  H.~Garcilazo, A.~Valcarce and T.~F.~Caram\'es,
  Eur.\ Phys.\ J.\ C {\bf 79}, 598 (2019). 

\bibitem{miyamoto18} 
  T.~Miyamoto {\it et al.},
  Nucl.\ Phys.\ A {\bf 971}, 113 (2018). 

\bibitem{Miyamoto:2018} 
  T.~Miyamoto [HAL QCD collaboration],
  PoS Hadron {\bf 2017}, 146 (2018).

\bibitem{Haidenbauer:2018} 
  J.~Haidenbauer and G.~Krein,
  Eur.\ Phys.\ J.\ A {\bf 54}, 199 (2018). 

\bibitem{Epelbaum:2003}
  E.~Epelbaum, U.-G.~Mei{\ss}ner, W.~Gl\"ockle,
  Nucl.\ Phys.\ A {\bf 714}, 535 (2003).

\bibitem{Petschauer:2013} 
  S.~Petschauer and N.~Kaiser,
  Nucl.\ Phys.\ A {\bf 916}, 1 (2013).

\bibitem{Gal:2016}
  A. Gal, E.V. Hungerford, D.J. Millener,
  Rev. Mod. Phys. {\bf 88}, 035004 (2016).

\bibitem{Reuber:1994} 
  A.~Reuber, K.~Holinde and J.~Speth,
  Nucl.\ Phys.\ A {\bf 570}, 543 (1994).

\bibitem{Haidenbauer:2005} 
  J.~Haidenbauer and U.-G.~Mei\ss ner,
  Phys.\ Rev.\ C {\bf 72}, 044005 (2005).

\bibitem{Meng:2019} 
  L.~Meng, B.~Wang and S.~L.~Zhu,
  arXiv:1912.09661 [nucl-th].


\bibitem{Huang:2013} 
  H.~Huang, J.~Ping and F.~Wang,
  Phys.\ Rev.\ C {\bf 87}, 034002 (2013).

\bibitem{Maeda:2018} 
  S.~Maeda, M.~Oka and Y.~R.~Liu,
  Phys.\ Rev.\ C {\bf 98}, 035203 (2018). 

\bibitem{Polinder} H. Polinder, J. Haidenbauer, and U.-G. Mei{\ss}ner,
 Nucl. Phys. A {\bf 779}, 244 (2006).

\bibitem{Haidenbauer:2013} 
  J.~Haidenbauer, S.~Petschauer, N.~Kaiser, U.-G.~Mei{\ss}ner, A.~Nogga and W.~Weise,
  Nucl.\ Phys.\ A {\bf 915}, 24 (2013).

\bibitem{Haidenbauer:2019} 
  J.~Haidenbauer, U.-G.~Mei{\ss}ner and A.~Nogga,
  arXiv:1906.11681 [nucl-th].
  
\bibitem{Petschauer:2020} 
  S.~Petschauer, J.~Haidenbauer, N.~Kaiser, U.-G.~Mei{\ss}ner and W.~Weise,
  Front.\ in Phys.\  {\bf 8}, 12 (2020).


\bibitem{Can:2016} 
  K.~U.~Can, G.~Erkol, M.~Oka and T.~T.~Takahashi,
  Phys.\ Lett.\ B {\bf 768}, 309 (2017).

\bibitem{Albertus:2005} 
  C.~Albertus, E.~Hernandez, J.~Nieves and J.~M.~Verde-Velasco,
  Phys.\ Rev.\ D {\bf 72}, 094022 (2005).

\bibitem{Alexandrou:2016} 
  C.~Alexandrou, K.~Hadjiyiannakou and C.~Kallidonis,
  Phys.\ Rev.\ D {\bf 94}, 034502 (2016).

\bibitem{PDG}
  M. Tanabashi et al. (Particle Data Group), Phys. Rev. D {\bf 98}, 030001 (2018).

\bibitem{Durr:2013} 
  S.~D\"urr {\it et al.} [Budapest-Marseille-Wuppertal collaboration],
  Phys.\ Rev.\ D {\bf 90}, 114504 (2014). 

\bibitem{Alexandrou:2014} 
  C.~Alexandrou, M.~Constantinou, K.~Hadjiyiannakou, K.~Jansen, C.~Kallidonis and G.~Koutsou,
  PoS LATTICE {\bf 2014}, 151 (2015).

\bibitem{Beane:2005} 
  S.~R.~Beane, P.~F.~Bedaque, A.~Parre\~no and M.~J.~Savage,
  Nucl.\ Phys.\ A {\bf 747}, 55 (2005). 
  
\bibitem{Epelbaum:2008}
  E.~Epelbaum, H.~W.~Hammer and U.-G.~Mei{\ss}ner,
  Rev.\ Mod.\ Phys.\  {\bf 81}, 1773 (2009).

\bibitem{Machleidt:2011} R.~Machleidt and D.~R. Entem,
  Phys. Rep. {\bf 503}, 1 (2011).
  

\bibitem{Haidenbauer:2015} 
  J.~Haidenbauer and U.~G.~Mei{\ss}ner,
  Nucl.\ Phys.\ A {\bf 936}, 29 (2015).
  
  
\bibitem{schulze97} H.-J. Schulze, M. Baldo, U. Lombardo, J. Cugnon and A. Lejeune,
 Phys. Rev. C {\bf 57}, 704 (1998). 

\bibitem{vidana00} I. Vida\~na, A. Polls, A. Ramos, M. Hjorth-Jensen and V. G. J. Stoks, 
 Phys. Rev. C {\bf 61}, 025802 (2002).

  \bibitem{Miyagawa:1995} 
  K.~Miyagawa, H.~Kamada, W.~Gl\"ockle and V.~G.~J.~Stoks,
  Phys.\ Rev.\ C {\bf 51}, 2905 (1995).
  
 \bibitem{Nogga:2002}
  A.~Nogga, H.~Kamada and W.~Gl\"ockle,
  Phys.\ Rev.\ Lett.\  {\bf 88}, 172501 (2002).
  
  \bibitem{Gibson:1994}
  B.~F.~Gibson, I.~R.~Afnan, J.~A.~Carlson and D.~R.~Lehman,
  Prog.\ Theor.\ Phys.\ Suppl.\  {\bf 117}, 339 (1994).
  
\bibitem{Nogga:2000uu} 
  A.~Nogga, H.~Kamada and W.~Gl\"ockle,
  Phys.\ Rev.\ Lett.\  {\bf 85}, 944 (2000).

\bibitem{Reinert:2018} 
  P.~Reinert, H.~Krebs and E.~Epelbaum,
  Eur.\ Phys.\ J.\ A {\bf 54}, 86 (2018).


\bibitem{borromeo92}  M. Borromeo, D. Bonatsos, H. M\"{u}ther, and A. Polls, Nucl. Phys. A {\bf 539}, 189 (1992).

\bibitem{morten94} M. Hjorth-Jensen, H. M\"{u}ther, and A. Polls,  Phys. Rev. C {\bf 50}, 501 (1994).

\bibitem{morten96} M. Hjorth-Jensen, A. Polls, A. Ramos,  and H. M\"{u}ther, Nucl. Phys. A {\bf 605}, 458 (1996).

\bibitem{vidana98} I. Vida\~na. A. Polls, A. Ramos, and M. Hjorth-Jensen, Nucl. Phys. A {\bf 644}, 201 (1998).

\bibitem{vidana17} I. Vida\~na, Nucl. Phys. A {\bf 958}, 48 (2017).

\bibitem{nogami70} Y. Nogami and E. Satoh, Nucl. Phys. B {\bf 19}, 93 (1970).

\bibitem{bodmer71} A. R. Bodmer and D. M. Rote, Nucl. Phys. A {\bf 169}, 1 (1971).

\bibitem{dabrowski73} J. Dabrowski, Phys. Lett. B {\bf 47}, 306 (1973).

\bibitem{tsushima03} K. Tsushima and F.~C. Khanna, Prog. Theor. Phys. Suppl. {\bf 149}, 160 (2003).

\bibitem{Haidenbauer:2019V} 
  J.~Haidenbauer and I.~Vida\~na,
  Eur. Phys. J. A {\bf 56}, 55 (2020).


\bibitem{Tsunemi:2008} 
 T. Tsunemi, presentation at ``Strangeness Nuclear Physics Experiments'',
Mito, Japan, March 5-6, 2008, \\
{\tt
http://nuclpart.kek.jp/NP08/presentations/\\
strangeness/pdf/Str13Tsunemi.pdf}

\bibitem{Steinheimer:2017} 
  J.~Steinheimer, A.~Botvina and M.~Bleicher,
  Phys.\ Rev.\ C {\bf 95}, 014911 (2017). 

\bibitem{Imai:2019} 
 K. Imai, presentation at ``J-PARC Symposium 2019'',
Tsukuba, Japan, September 23-27, 2019, 
{\tt
https://conference-indico.kek.jp/indico/\\
event/91/session/45/contribution/122/\\
material/slides/0.pdf}

\bibitem{Cho:2017} 
  S.~Cho {\it et al.} [ExHIC Collaboration],
  Prog.\ Part.\ Nucl.\ Phys.\  {\bf 95}, 279 (2017). 

\bibitem{Braun:2018} 
  P.~Braun-Munzinger and B.~D\"onigus,
  Nucl.\ Phys.\ A {\bf 987}, 144 (2019). 
  
  \bibitem{Donigus:2019}
 ALICE Collaboration, {\it Letter of Intent}, 
 {\tt
  https://cds.cern.ch/record/2703140}
  
  \bibitem{Agakishiev:2011} 
  H.~Agakishiev {\it et al.} [STAR collaboration],
  Nature {\bf 473}, 353 (2011)
  Erratum: [Nature {\bf 475}, 412 (2011)].
  
  \bibitem{TheSTAR:2016} 
  [STAR collaboration and CBM eTOF group],
  arXiv:1609.05102 [nucl-ex].


\bibitem{Lu:2017}
  J.-X.~Lu, L.-S.~Geng and M.~Pav\'on Valderrama,
  Phys.\ Rev.\ D {\bf 99}, 074026 (2019). 

\bibitem{Baru:2011} 
  V.~Baru, A.~A.~Filin, C.~Hanhart, Y.~S.~Kalashnikova, A.~E.~Kudryavtsev and A.~V.~Nefediev,
  Phys.\ Rev.\ D {\bf 84}, 074029 (2011)

\bibitem{Baru:2013} 
  V.~Baru, E.~Epelbaum, A.~A.~Filin, C.~Hanhart, U.-G.~Mei{\ss}ner and A.~V.~Nefediev,
  Phys.\ Lett.\ B {\bf 726}, 537 (2013). 

\bibitem{Baru:2015D} 
  V.~Baru, E.~Epelbaum, A.~A.~Filin, J.~Gegelia and A.~V.~Nefediev,
  Phys.\ Rev.\ D {\bf 92}, 114016 (2015). 

  \end{thebibliography}
  \end{document}